\documentclass[final, 3p, times, 10pt]{elsarticle}

\usepackage{amsmath}
\usepackage{amssymb}
\usepackage{color}
\DeclareMathAlphabet{\altmathcal}{OMS}{cmsy}{m}{n}
\usepackage{mathptmx}
\usepackage{color}
\usepackage{multirow}
\usepackage{aas_macros}
\usepackage{ulem}
\usepackage{graphicx}
\usepackage{subfigure}
\usepackage{xcolor}
\usepackage{color, soul}
\usepackage{enumitem}

\journal{Journal of Computational Physics}

\begin{document}


\newcommand{\av}[1]{\left<{#1}\right>}
\newcommand{\DS}{\displaystyle}
\newcommand{\HALF}{\frac{1}{2}}
\newcommand{\hvec}[1]{\hat{\mathbf{#1}}}
\newcommand{\pd}[2]{\frac{\partial #1}{\partial #2}}
\newcommand{\quotes}[1]{``#1''}
\newcommand{\tens}[1]{\mathsf{#1}}
\renewcommand{\vec}[1]{\mathbf{#1}}

\newcommand{\cE}{\altmathcal{E}}
\newcommand{\cF}{\altmathcal{F}}
\newcommand{\cR}{\altmathcal{R}}
\newcommand{\cU}{\altmathcal{U}}
\newcommand{\cV}{\altmathcal{V}}

\newcommand{\cc}{ {\boldsymbol{c}} }
\newcommand{\cf}{ f }


\newcommand{\xc}{ {\mathbf{x}_c} }
\newcommand{\yc}{ {\mathbf{y}_c} }
\newcommand{\zc}{ {\mathbf{z}_c} }

\newcommand{\xf}{ {\mathbf{x}_f} }
\newcommand{\yf}{ {\mathbf{y}_f} }
\newcommand{\zf}{ {\mathbf{z}_f} }

\newcommand{\xe}{ {\mathbf{x}_e} }
\newcommand{\ye}{ {\mathbf{y}_e} }
\newcommand{\ze}{ {\mathbf{z}_e} }


\newcommand{\favx}[1]{\hat{#1}_{x,\xf}}
\newcommand{\favy}[1]{\hat{#1}_{y,\yf}}
\newcommand{\favz}[1]{\hat{#1}_{z,\zf}}

\newcommand{\lavx}[1]{\bar{#1}_{x,\xe}}
\newcommand{\lavy}[1]{\bar{#1}_{y,\ye}}
\newcommand{\lavz}[1]{\bar{#1}_{z,\ze}}


\newcommand{\RED}{\color{red}}
\newcommand{\BLACK}{\color{black}}
\newcommand{\BLUE}{\color{blue}}
\newcommand{\MAGENTA}{\color{magenta}}
\newcommand{\GREEN}{\color{olive}}


\newcommand{\rvec}{\mathrm {\mathbf {r}}} 


\newcommand{\refig}[1]{Fig.~\ref{#1}}
\newcommand{\refeq}[1]{Eq.~(\ref{#1})}

\begin{frontmatter}



\title{A $4^{\rm th}$-order accurate finite volume method for ideal classical and special relativistic MHD based on pointwise reconstructions}


\author[address1]{V. Berta}
\author[address1]{A. Mignone}
\author[address1,address2]{M. Bugli}
\author[address3]{G. Mattia}

\address[address1]
{Dipartimento di Fisica, Universit\`a di Torino, via Pietro Giuria 1, 10125 Torino, Italy}
\address[address2]
{Universit\'e Paris-Saclay, Universit\'e Paris Cit\'e, CEA, CNRS, AIM, 91191, Gif-sur-Yvette, France}
\address[address3]
{INFN, Sezione di Firenze, Via G. Sansone 1, I-50019 Sesto Fiorentino, Italy}

%

\begin{abstract}
We present a novel implementation of a genuinely $4^{\rm th}$-order accurate finite volume scheme for multidimensional classical and special relativistic magnetohydrodynamics (MHD) based on the constrained transport (CT) formalism. 
The scheme introduces several novel aspects when compared to its predecessors yielding a more efficient computational tool.
Among the most relevant ones, our scheme exploits pointwise to pointwise reconstructions (rather than one-dimensional finite volume ones), employs the generic upwind constrained transport averaging and sophisticated limiting strategies that include both a discontinuity detector and an order reduction procedure.
Selected numerical benchmarks demonstrate the accuracy and robustness of the method.
\end{abstract}


\end{frontmatter}


\section{Introduction}
%
%
%

The modeling of astrophysical plasmas is nowadays demanding for more efficient and accurate numerical schemes for solving the equations of classical and special relativistic magnetohydrodynamics (MHD, RMHD).
Physical interest is increasingly directed towards the investigation of highly nonlinear flows featuring both smooth and discontinuous solutions, with a great deal of attention being devoted to enhance the accuracy of the underlying numerical methods while retaining stability and computational efficiency.

Traditionally, in the context of finite volume (FV) or finite difference (FD) schemes, a great variety of $2^{\rm nd}$-order methods have been developed to solve the MHD and RMHD equations (e.g., \cite{Falle_Komissarov1998, Balsara1998, Dai_Woodward1998, Powell_etal1999, Balsara_Spicer1999, Toth2000, Balsara2004, Torrilhon2005, Gardiner_Stone2005, Fromang2006, Mignone_Bodo2006, Rossmanith2006, Toth_Gabor_2006, Lee_Deane2009, Mignone_etal2010, Meyer_2012, Mignone_etal2012, Mignone_DelZanna2021}). 
However, in the last decade several efforts have been made to design higher than $2^{\rm nd}$-order numerical methods both for FD (e.g., \citep{Londrillo_DelZanna2000, DelZanna_etal2007, Mignone_etal2010, Christilieb_2014, Chen_2016, Minoshima_etal2019, Reyes_2019}) as well as for FV (e.g., \citep{ Shu2009, Balsara2009, ColellaDorr_2011, Corquodale_Colella2011, Helzel_etal2011, Susanto_2013, Balsara2017, Lee_2017}) approaches, leading to the dawn of high-order numerical codes (e.g. \citep{Li_2010, Nunez_2016a, Nunez_2016b, Felker_Stone2018, Matsumoto_etal2019, Verma_etal2019}) in plasma physics and computational fluid-dynamics.
Most $4^{\rm th}$-order schemes, in fact, can reach unprecedented performances overstepping the limits of traditional $2^{\rm nd}$-order frameworks because of their intrinsic lower dissipation properties.
In addition, smooth solutions are improved at much faster rates, yielding enhanced convergence and henceforth a computational efficiency that increases with the dimensionality of the problem and with resolution.

In this paper we present a genuine $4^{\rm th}$-order accurate finite volume method for the numerical solution of the ideal classical and special relativistic MHD equations.
Such method draws on the work originally introduced by McCorquodale \& Colella \citep{Corquodale_Colella2011} and then refined by Felker \& Stone \citep{Felker_Stone2018}, but with a number of important improvements and differences.
In fact, our numerical scheme is introducing, for the first time in the literature of FV schemes, a version of the MP5 \citep{Suresh_Huynh1997} and the WENOZ \citep{Jiang_Shu1996, Borges_WENOZ2008, Castro_2011} spatial reconstruction algorithms based on pointwise values, which are instead traditionally used in FD schemes. 
As we shall demonstrate, pointwise reconstructions lower the algorithmic complexity of the overall scheme by retaining the $4^{\rm th}$-order accuracy with fewer operations per step and without even enlarging the stencil size.
Moreover, this new feature reduces the number of Riemann problems to be solved per direction, leading to a more efficient and cost-effective scheme.
Concurrently, robustness is ensured by means of a local discontinuity detector to distinguish smooth solutions from discontinuous ones when determining the function local point value from its cell average.
Regarding temporal integration, we employ a semi-discrete approach based on a five-stage $4^{\rm th}$-order explicit strong stability preserving Runge-Kutta method (eSSPRK(5,4), \citep{Isherwood_RK42018, Spiteri_Ruuth2002}).
The solenoidal condition of the magnetic field is fulfilled by means of a high-order formulation of the generalized upwind constrained transport (UCT, \citep{Evans_Hawley1988, Dai_Woodward1998, Balsara_Spicer1999, Ryu_etal1998, Londrillo_DelZanna2004}) algorithm, which ensures a divergence-free magnetic field up to machine accuracy, and works well with any generic upwind average \citep{Mignone_DelZanna2021}.

The paper is structured as follow: in \S 2 we will briefly describe the MHD and RMHD equations and how they are discretized in a general FV framework together with the notation used throughout this work. 
In \S 3 an extensive description of the novel numerical method will be provided. 
In \S 4 we will illustrate our method to control the solenoidal condition. 
Numerical tests and conclusions will be given, respectively, in \S 5 and \S 6.

\section{The ideal MHD and RMHD equations}
\label{sec:equations}
%
%
%

The ideal MHD equations can be split out in two coupled sub-systems \citep{Londrillo_DelZanna2000, Londrillo_DelZanna2004, Mignone_DelZanna2021}, the first one being a time-dependent hyperbolic system for the evolution of the conservative flow variables (the mass density $\rho$, the momentum density $\rho\vec{v}$, and the energy density $\cE$). 
This sub-system can be written in compact form by introducing a state vector of conservative quantities, $U$, and a rank-2 tensor, $\tens{F}$, whose rows represent the fluxes relative to each component of $U$
\begin{equation}\label{eq:FV}
  \pd{U}{t} + \nabla\cdot\tens{F} = 0 \, ,
\end{equation}
where
\begin{equation} \label{eq:MHD}
   U = \left(\begin{array}{l}
     \rho          \\ \noalign{\medskip}
     \rho \vec{v}  \\ \noalign{\medskip}
     \cE      
  \end{array}\right) \,,\quad
  \tens{F} = \left(\begin{array}{c}
     \rho \vec{v}                                         \\ \noalign{\medskip}
     \rho \vec{v}\vec{v} - \vec{B}\vec{B} + \tens{I}p_t   \\ \noalign{\medskip}
     (\cE + p_t)\vec{v} - (\vec{v}\cdot\vec{B})\vec{B}
  \end{array}\right)^\intercal \,.\quad
  \vspace{0.3 cm}
\end{equation}
In \refeq{eq:MHD}, $\vec{v} = (v_x,\, v_y,\, v_z)$ is the fluid velocity, $\vec{B} = (B_x,\, B_y,\, B_z)$ is the magnetic field, $p_t = p+B^2/2$ is the total pressure expressed as the sum of the thermal and magnetic pressure, and $\tens{I}$ is the identity matrix. 
The total energy density is composed by a kinetic, a thermal and a magnetic term
\begin{equation}
   \cE = \frac{1}{2}\rho\vec{v}^2 + \frac{p}{\Gamma-1} + \frac{\vec{B}^2}{2} \,,
\end{equation}
while $\Gamma$ is the specific heat ratio for an adiabatic equation of state.
On the other hand, the second sub-system consists of the induction equation for the evolution of the magnetic field
\begin{equation}\label{eq:induction}
  \frac{\partial\vec{B}}{\partial t} + c\nabla\times\vec{E}=0.
\end{equation}
Due to the presence of the curl operator instead of the divergence, the latter equation cannot be considered a simple extension of the HD subsystem \citep{Londrillo_DelZanna2000}. 
Additionally, in the ideal MHD framework, the electric field $\vec{E}$ is a function of the fluid velocity $\vec{v}$ and the magnetic field $\vec{B}$
\begin{equation}
c\vec{E} = - \vec{v} \times \vec{B} \,,
\end{equation}
where $c$ is the speed of light.

Analytically, the commutativity of spatial derivatives endows the induction equation with the solenoidal condition for the magnetic field
\begin{equation} \label{eq:divB}
    \nabla\cdot\vec{B} = 0.
\end{equation}
This stationary condition, once satisﬁed at $t = 0$, is analytically preserved at $t > 0$. 
On the other hand, since numerical derivatives do not commute, the condition expressed in \refeq{eq:divB} is not automatically fulfilled by numerical schemes, leading to the rising of magnetic monopoles which cause unphysical plasma transport and dissipation of momentum and energy in ideal frameworks \citep{Brackbill_Barnes1980}. 
For these reasons, HD numerical schemes cannot be straightforwardly extended to MHD systems without introducing some corrections to numerically constraint \refeq{eq:divB} to be satisfied.

Similarly to the MHD case, the ideal RMHD equations can be expressed in the form given by \refeq{eq:FV} with
\begin{equation} \label{eq:RMHD}
   U = \left(\begin{array}{l}
     D          \\ \noalign{\medskip}
     \vec{m}  \\ \noalign{\medskip}
     \cE      
  \end{array}\right) \,,\quad
  \tens{F} = \left(\begin{array}{c}
     D \vec{v}                                         \\ \noalign{\medskip}
     w_t\gamma^2 \vec{v}\vec{v} - \vec{b}\vec{b} + \tens{I}p_t   \\ \noalign{\medskip}
     \vec{m}
  \end{array}\right)^\intercal \,,\quad
  \vspace{0.3 cm}
\end{equation}
where $D=\gamma\rho$, $\vec{m}=(\rho h\gamma^2+|\vec{B}|^2)\vec{v}-(\vec{v}\cdot\vec{B})\vec{B}$, and $\cE=\rho h\gamma^2-p+|\vec{B}|^2/2+(|\vec{v}|^2|\vec{B}|^2-(\vec{v}\cdot\vec{B})^2)/2$ are respectively the relativistic mass density, momentum density, and energy density, while $w_t=\rho h+|\vec{B}|^2/\gamma^2+(\vec{v}\cdot\vec{B})^2$ and $\vec{b}=\vec{B}/\gamma+\gamma(\vec{v}\cdot\vec{B})\vec{v}$.
In the previous expressions we introduced the specific enthalpy $h$, the Lorentz factor $\gamma=(1-|\vec{v}|^2)^{-1/2}$ and we set the speed of light $c$ to unity.
In addition, the same induction equation as in the MHD regime needs to be added to \refeq{eq:RMHD}, thus requiring the same numerical scheme to control the divergence-free condition shown in \refeq{eq:divB}.

\subsection{Notations}
\label{sec:notations}
%

We employ a Cartesian coordinate system with unit vectors $\hvec{e}_x=(1,0,0)$, $\hvec{e}_y=(0,1,0)$, and $\hvec{e}_z=(0,0,1)$ and mantain the same notation formalism adopted by Mignone \& Del Zanna \cite{Mignone_DelZanna2021}.
In each direction the coordinate spacing $\Delta x$, $\Delta y$, and $\Delta z$ is uniform. 
Computational cells are centered at $(x_i,\, y_j,\, z_k)$ and delimited by the six interfaces orthogonal to the coordinate axis centered, respectively, at $(x_{i\pm\HALF},\, y_j,\, z_k)$, $(x_i,\, y_{j\pm\HALF},\, z_k)$, and $(x_i,\, y_j,\, z_{k\pm\HALF})$.

According to Gauss' theorem, the conserved quantities are evolved as volume averages $\av{U}_{\cc}$ over the cell volume
\begin{equation} \label{eq:avg}
    \av{U}_{\cc} \equiv \frac{1}{\Delta x \Delta y \Delta z} \int U(x,y,z,t) \,dxdydz \,,
\end{equation}
where the $\cc$ subscript is a shorthand notation for $(i,j,k)$, by means of the neat difference of the surface-averaged representation of the fluxes at zone interfaces:
\begin{equation} \label{eq:fluxes}
  \begin{array}{l}
  \DS \favx{F} \equiv \frac{1}{\Delta y\Delta z}
      \int \hvec{e}_x\cdot\tens{F}\big(U(x_{i+\HALF},y,z,t)\big)  \,dy\,dz \,,
  \\ \noalign{\medskip}
  \DS \favy{F} \equiv \frac{1}{\Delta x\Delta z}
      \int \hvec{e}_y\cdot\tens{F}\big(U(x,y_{j+\HALF},z,t)\big)  \,dz\,dx \,,
  \\ \noalign{\medskip}
  \DS \favz{F} \equiv \frac{1}{\Delta x\Delta y}
      \int \hvec{e}_z\cdot\tens{F}\big(U(x,y,z_{k+\HALF},t)\big)  \,dx\,dy \,.
  \end{array}
\end{equation}
In the finite-volume fashion, a semi-discrete method of lines approximates the partial differential equation (PDE) associated with \refeq{eq:FV}, yielding an ordinary differential equation (ODE) in the time variable 
\begin{equation} \label{eq:ode}
    \DS \frac{d \av{U}_{\cc}}{dt} = - \left(
        \frac{\favx{F} - \hat{F}_{x,\xf-\hvec{e}_x}}{\Delta x} 
      + \frac{\favy{F} - \hat{F}_{y,\yf-\hvec{e}_y}}{\Delta y} 
      + \frac{\favz{F} - \hat{F}_{z,\zf-\hvec{e}_z}}{\Delta z} 
        \right) \, .
\end{equation}

Eqns.~(\ref{eq:avg}-\ref{eq:ode}) frame the discretization for zone-centered variables.
In constrained transport MHD instead, the magnetic field update relies on a discrete version of Stoke's theorem \citep{Yee1966, Brecht_etal1981, Evans_Hawley1988, Balsara_Spicer1999}, where the magnetic field is treated as a staggered primary variable. 
In this representation, each component of the magnetic field 
\begin{equation}
  \vec{\hat{B}}_{\cf} \equiv 
    \left( \begin{array}{l}
              \hat{B}_{x,i+\HALF,j,k}  \\ \noalign{\medskip}
              \hat{B}_{y,i,j+\HALF,k}  \\ \noalign{\medskip}
              \hat{B}_{z,i,j,k+\HALF}     \end{array}\right)
    =  \left( \begin{array}{l}
              \favx{B}  \\ \noalign{\medskip}
              \favy{B}  \\ \noalign{\medskip}
              \favz{B}     \end{array}\right) \,
\end{equation}
is evolved as an area-weighted average on the face orthogonal to the field component.
The subscripts $\xf$, $\yf$ and $\zf$ identify the location of the faces, i.e., $\xf \equiv (i+\HALF, j, k)$, $\yf \equiv (i, j+\HALF, k)$, and $\zf \equiv (i, j, k+\HALF)$.

Furthermore, \refeq{eq:induction} prescribes that the variation in time of the magnetic flux is equal to the line integral of the electric field $\bar{\vec{E}}_{e}$ over zone edges, namely, the electromotive force (EMF)
\begin{equation}
  \bar{\vec{E}}_{e} \equiv 
    \left( \begin{array}{l}
              \bar{E}_{x,i,j+\HALF,k+\HALF}  \\ \noalign{\medskip}
              \bar{E}_{y,i+\HALF,j,k+\HALF}  \\ \noalign{\medskip}
              \bar{E}_{z,i+\HALF,j+\HALF,k}     \end{array}\right)
  =  \left(\begin{array}{l}
              \lavx{E}  \\ \noalign{\medskip}
              \lavy{E}  \\ \noalign{\medskip}
              \lavz{E}     \end{array}\right)  \,,
\end{equation}
where:
\begin{equation}
  \begin{array}{l}
  \DS \lavx{E} \equiv \frac{1}{\Delta x}
      \int E_x(x,y_{j+\HALF},z_{k+\HALF},t) \,dx\,,  \\ \noalign{\medskip}
  \DS \lavy{E} \equiv \frac{1}{\Delta y}
      \int E_y(x_{i+\HALF},y,z_{k+\HALF},t) \,dy\,,  \\ \noalign{\medskip}
  \DS \lavz{E} \equiv \frac{1}{\Delta z}
      \int E_z(x_{i+\HALF},y_{j+\HALF},z,t) \,dz\,.
  \end{array}
\end{equation}
Similarly to the magnetic field, the edge-centered electric field positions are labeled as $\xe \equiv (i, j+\HALF, k+\HALF)$, $\ye \equiv (i+\HALF, j, k+\HALF)$, and $\ze \equiv (i+\HALF, j+\HALF, k)$.
Given all these elements, it is possible to retrieve the discretized version of Stokes theorem
\begin{equation} \label{eq:stokes}
 \begin{array}{lcl}
  \DS \frac{d  \favx{B}}{dt} & = & \DS - 
       \left(  \frac{\bar{E}_{z,\ze} - \bar{E}_{z,\ze-\hvec{e}_y}}
                    {\Delta y}
              -\frac{\bar{E}_{y,\ye} - \bar{E}_{y,\ye-\hvec{e}_z}}
                    {\Delta z}
       \right)  \,, \\ \noalign{\medskip}         
  \DS \frac{d  \favy{B}}{dt} & = & \DS - 
       \left(  \frac{\bar{E}_{x,\xe} - \bar{E}_{x,\xe-\hvec{e}_z}}
                    {\Delta z}
              -\frac{\bar{E}_{z,\ze} - \bar{E}_{z,\ze-\hvec{e}_x}}
                    {\Delta x}
       \right)  \,, \\ \noalign{\medskip}         
  \DS \frac{d  \favz{B}}{dt} & = & \DS - 
       \left(  \frac{\bar{E}_{y,\ye} - \bar{E}_{y,\ye-\hvec{e}_x}}
                    {\Delta x}
              -\frac{\bar{E}_{x,\xe} - \bar{E}_{x,\xe-\hvec{e}_y}}
                    {\Delta y}  
       \right)\,,
\end{array}
\end{equation}
where we set $c=1$, thus verifying
\begin{equation} \label{eq:divbmh}
 \frac{d}{dt} \left( 
           \frac{\hat{B}_{x,\xf} - \hat{B}_{x,\xf-\hvec{e}_x}}
                {\Delta x}
         + \frac{\hat{B}_{y,\yf} - \hat{B}_{y,\yf-\hvec{e}_y}}
                {\Delta y}
         + \frac{\hat{B}_{z,\zf} - \hat{B}_{z,\zf-\hvec{e}_z}}
                {\Delta z}
         \right) = 0 \, ,
\end{equation}
which is valid exactly and guaranties the fulfillment of the solenoidal condition for an initially divergence-free magnetic field.
Notice that no approximation has been made in the treatment so far.
To summarize our notation convention, given the generic physical quantity $Q$ we indicate with $\av{Q}_{\cc}$ its volume average.
We use $\hat{Q}_{\xf, \yf, \zf}$ and $\bar{Q}_{\xe, \ye, \ze}$ to refer to the face and edge averages, respectively.
Similarly, the notations $Q_{\cc}$, $Q_{\xf, \yf, \zf}$, and $Q_{\xe, \ye, \ze}$ refer to the pointwise values at the center of the cell, face, and edge, respectively.

\section{The $4^{\rm th}$-order accurate method}
\label{sec:ho_scheme}
%
%
%

In this section we introduce the main features of our $4^{\rm th}$-order accurate numerical scheme.
Finite volume Godunov-type methods evolve the cell-averaged conservative variables (e.g., ~$\av{U}_\cc = \{\av{\rho}_\cc, \av{\rho\vec{v}}_\cc, \av{\cE}_\cc$\}, while the representation of $\Vec{B}$ depends on the choice of the algorithm chosen to control the solenoidal condition and may be either a cell-centered or a staggered quantity) and typically consist of three main stages: a reconstruction step at cell faces, flux computation at zone interfaces via Riemann solvers, and time integration.
At $2^{\rm nd}$-order accuracy, point values can be interchanged with volume or surface averages, so that the transformation between conservative and primitive variables (e.g., ~$V_c = \{\rho, \vec{v}, p\}$ same arguments as before hold for $\Vec{B}$) can be simply defined as $V_\cc \approx \av{V}_{\cc} \approx {\mathcal V} (\av{U}_\cc)$.
This property, however, does no longer hold for higher order schemes, where one has to distinguish the cell-centered point value from its volume average and, likewise, the face-centered flux from its surface average. 
On top of that, one-dimensional FV reconstruction schemes have to be reformulated in order to obtain $4^{\rm th}$-order accurate left and right states at zone interfaces. 
These states are best obtained using primitive or characteristic variables in order to avoid unwanted numerical oscillations. 
These steps are described in the following sections.

\subsection{Point value recovery of primitive variables}
\label{sec:PointValue}
%

Before reconstruction of primitive variables can take place, their values must become available at cell centers.
Since conservative schemes evolve the volume average of conservative quantities $\av{Q}_\cc$, their local point value $Q_\cc$ has to be obtained first.
To $4^{\rm th}$-order accuracy, point values can be retrieved from cell averages using the relation
\begin{equation}\label{eq:v2p}
    Q_\cc = \DS \av{Q}_\cc - \frac{\Delta\av{Q}_\cc}{24} + O(h^4) \, ,
\end{equation}
and conversely
\begin{equation}\label{eq:p2v}
    \av{Q}_\cc = Q_\cc + \frac{\Delta Q_\cc}{24} + O(h^4) \,, 
\end{equation}
where \refeq{eq:p2v} is effectively a high-order Simpson-like quadrature rule.
In Eqns. (\ref{eq:v2p}) and (\ref{eq:p2v}) $Q$ is a component of $U_\cc$, while $\Delta$ is the Laplacian operator first introduced in the context of high-order methods by McCorquodale \& Colella \citep{Corquodale_Colella2011}, i.e.
\begin{equation} \label{eq:laplacian}
  \Delta Q_{\cc}  \equiv 
     \Delta^{x}Q_\cc +
     \Delta^{y}Q_\cc +
     \Delta^{z}Q_\cc \, ,
\end{equation}
with components given by:
\begin{equation} \label{eq:laplacian1D}
  \Delta^{x} Q_{\cc}  \equiv  
  \DS \left (Q_{\cc - \hvec{e}_x} - 2Q_{\cc} + 
             Q_{\cc + \hvec{e}_x} \right) \, ,
  \,\quad
  \Delta^{y} Q_{\cc}  \equiv  
  \DS \left ( Q_{\cc - \hvec{e}_y} - 2Q_{\cc}
             +Q_{\cc + \hvec{e}_y}\right) \, ,
  \,\quad
  \Delta^{z} Q_{\cc}  \equiv  
  \DS \left (Q_{\cc - \hvec{e}_z} - 2Q_{\cc}
             +Q_{\cc + \hvec{e}_z}\right)  \,.
\end{equation}
The interested reader may refer to \ref{app:laplacians} for a full derivation of Eqns. (\ref{eq:v2p}) and (\ref{eq:p2v}).

Similarly, high-order transformations need to be applied also in the conversion between point values and face- or edge-averaged quantities, by means of the corresponding 2D or 1D operators.
At zone faces, for instance, we use transverse Laplacian operators defined by:
\begin{equation}
    \Delta^x_\bot Q_{\xf} = \Delta^y Q_\xf + \Delta^z Q_\xf \, , \quad
    \Delta^y_\bot Q_{\yf} = \Delta^x Q_\yf + \Delta^z Q_\yf \, , \quad
    \Delta^z_\bot Q_{\zf} = \Delta^x Q_\zf + \Delta^y Q_\zf \, .
\end{equation}
This is employed, for example, to recover the face-centered point value of the magnetic field from its surface average:
\begin{equation}\label{eq:v2p_B}
  B_{x,\xf} = \favx{B} - \frac{\Delta^{x}_\bot \favx{B}}{24} , \quad  
  B_{y,\yf} = \favy{B} - \frac{\Delta^{y}_\bot \favy{B}}{24} , \quad  
  B_{z,\zf} = \favz{B} - \frac{\Delta^{z}_\bot \favz{B}}{24} \, .
\end{equation}
Following Felker \& Stone \citep{Felker_Stone2018}, the pointwise value of the magnetic field at the cell center is obtained using an unlimited high-order interpolation, i.e.:
\begin{equation}\label{eq:Bcp}
 \begin{array}{l}
  \DS B_{x, c} = - \frac{1}{16}(B_{x,\xf+\hvec{e}_x} + B_{x,\xf-2\hvec{e}_x}) 
                + \frac{9}{16}(B_{x,\xf} + B_{x,\xf-\hvec{e}_x}) \, ,
  \\ \noalign{\medskip}
  \DS B_{y, c} = - \frac{1}{16}(B_{y,\yf+\hvec{e}_y} + B_{y,\yf-2\hvec{e}_y}) 
                + \frac{9}{16}(B_{y,\yf} + B_{y,\yf-\hvec{e}_y}) \, ,
  \\ \noalign{\medskip}
  \DS B_{z, c} = - \frac{1}{16}(B_{z,\zf+\hvec{e}_z} + B_{z,\zf-2\hvec{e}_z}) 
                + \frac{9}{16}(B_{z,\zf} + B_{z,\zf-\hvec{e}_z}) \, .
  \end{array}
\end{equation}
Once the pointwise values of the conservative variables are obtained by means of Eqns. (\ref{eq:v2p}) and (\ref{eq:Bcp}), the conversion to primitive variables takes place.

\subsection{The pointwise MP5 and WENOZ spatial reconstruction algorithms}
\label{sec:point_rec}
%

Here we introduce for the first time in a FV framework, to the extent of our knowledge, reconstruction schemes that operate directly on the function point values rather than on one-dimensional average quantities as traditionally done by previous investigations (see, e.g., \citep{Jiang_Shu1996, Suresh_Huynh1997, Titarev_Toro_2004, Corquodale_Colella2011, Balsara2017, Matsumoto_etal2019}). 
Compared to other FV $4^{\rm th}$-order methods, such as the one proposed by Felker \& Stone \citep{Felker_Stone2018}, we do not need to convert the average value $\av{U}_{\cc}$ into the mapped value $\mathcal{V}(\av{U_{\cc}})$, nor to compute from it the average value $\av{V}_{\cc}$ before reconstructing, thus resulting in a more efficient and cost-effective scheme.
In addition, the advantages of a point value reconstruction include that left and right states after reconstruction can be directly fed into the Riemann solver, avoiding the computation of interface-averaged Riemann problems.
A similar feature can also be found in the scheme presented by N\'u\~nez-de La Rosa \& Munz \cite{Nunez_2016a}, as they reconstruct point values of the conservative quantities at the interfaces starting from cell-averaged values via a multi-dimensional Gaussian quadrature rule \cite{Titarev_Toro_2004}.
However, the approach from N\'u\~nez-de La Rosa \& Munz is meant to reconstruct either conservative or characteristic quantities (whose volume-averaged values are the only available at the beginning of the integration step), while in our implementation we can either reconstruct primitive or characteristic variables (which in both approaches increase the stability of the scheme). 
More importantly, the high-order reconstruction based on the approach by \cite{Titarev_Toro_2004} leads to a number of Riemann problems at each cell's interface equal to the number of Gaussian points used during the integration (e.g. $4$ in the case of a three-dimensional domain when using $2$ Gaussian points).
Our scheme, on the other hand, always produces one Riemann problem per cell's interface.  

The monotonicity preserving (MP5) scheme of Suresh \& Huynh \citep{Suresh_Huynh1997} provides a $5^{\rm th}$-order accurate polynomial reconstruction at cell faces and, when necessary, limits the reconstructed value to preserve monotonicity near discontinuities while ensuring accuracy in smooth regions. 
While in the original formulation the unlimited interface value is retrieved by one-dimensional cell averages, our scheme replaces its definition by using point values in the same fashion as FD schemes (see, e.g., Appendix A.2 of Del Zanna et al. \cite{DelZanna_etal2007} for a brief summary).
The unlimited value at $i + \HALF$ is now given by
\begin{equation} \label{eq:mp5}
    \left.\ P(x)\right|_{i+\HALF} \equiv V^L_{i + \HALF} = 
    \frac{1}{128}\left(3V_{i-2} -20V_{i-1} +90V_i + 60V_{i+1} -5V_{i+2}\right) \, .
    \vspace{0.2 cm}
\end{equation}
where $\{ V_{i\pm2}, V_{i\pm1}, V_{i} \}$ are now  primitive quantities evaluated at cell-centers.
The scheme employs, as in the original case, a five-point stencil in order to distinguish between local extrema and a genuine O(1) discontinuity.
The MP5 limiting procedure remains unchanged as monotonicity-preserving bounds are equally well expressed in terms of point values.
We refer the reader to the original work of Suresh \& Huynh \cite{Suresh_Huynh1997} or to Appendix 2 of Del Zanna et al. \cite{DelZanna_etal2007}.

Likewise, we also introduce a pointwise version of the $5^{\rm th}$-order weighted essentially non-oscillatory (WENO, \citep{Jiang_Shu1996}) scheme in the version proposed by Borges et al. (WENOZ, \citep{Borges_WENOZ2008}), which attains better spatial resolution with reduced numerical dissipation at a modest computational cost.
The same interface value of \refeq{eq:mp5} at $x=x_{i+\HALF}$ is computed as the convex combination of $3^{\rm rd}$-order accurate interface values built on the three possible three-point sub-stencils $\{i-2,i-1,i\}$, $\{i-1,i,i+1\}$, and $\{i,i+1,i+2\}$, i.e.
\begin{equation}\label{eq:wenoz}
    \DS V^L_{i+\HALF} = \DS
    \omega_0 \frac{3V_{i-2} - 10V_{i-1} + 15V_i}{8}
  + \omega_1 \frac{-V_{i-1} +  6V_{i}   + 3V_{i+1}}{8}
  + \omega_2 \frac{3V_{i}   +  6V_{i+1} - V_{i+2}}{8} \, ,
\end{equation}
where, again, $\{ V_{i\pm2}, V_{i\pm1}, V_{i} \}$ are point values.
The weights $\omega_l$, for $l=\{0,1,2\}$, are defined as
\begin{equation}
    \omega_l = \frac{\alpha_l}{\sum_{m} \alpha_m}, \hspace{0.5cm} \alpha_l = d_l\left(1+ \frac{|\beta_0-\beta_2|}{\beta_l+\epsilon}\right) \, .
\end{equation}
Here $d_l$ denotes the optimal weights that reckon the $5^{\rm th}$-order accurate approximation of \refeq{eq:mp5}, while $\epsilon = 10^{-40}$ is a small number which avoids division by zero.
In our new framework, the optimal weights $\{d_0 = 1/16, d_1 = 5/8, d_2 = 5/16 \}$ can be determined from the uniqueness of the interpolating polynomial with $d_0+d_1+d_2=1$.
%
%
On the other hand, the smoothness indicators $\beta_l$ estimate the regularity of the polynomial approximation and, for a reconstruction based on point values, have been found to be the same as in Borges \citep{Borges_WENOZ2008}:
\begin{equation}
  \begin{aligned}
    \beta_0 &= \frac{13}{12}(V_{i-2} - 2V_{i-1} + V_i)^2 
             + \frac{1}{4}(V_{i-2} - 4V_{i-1} + 3V_{i})^2 \, ,
             \\ \noalign{\smallskip}
    \beta_1 &= \frac{13}{12}(V_{i-1} - 2V_{i} + V_{i+1})^2 
             + \frac{1}{4}(V_{i-1} - V_{i+1})^2  \, ,
             \\ \noalign{\smallskip}
    \beta_2 &= \frac{13}{12}(V_{i} - 2V_{i+1} + V_{i+2})^2 
             + \frac{1}{4}(3V_{i} - 4V_{i+1} + V_{i+2})^2 \, .
  \end{aligned}
\end{equation}
Note that the right state, $V_{i-\HALF}^R$, is obtained similarly by reversing the index order (e.g. $i-1 \to i+1$, etc).

We will also occasionally employ the $3^{\rm rd}$-order WENO scheme (here denoted as WENO3) \cite{Jiang_Shu1996} in the improved version of Yamaleev \& Carpenter \cite{Yamaleev_Carpenter2009}. 
The point value approach leads to the following expression for the left interface value
\begin{equation}
  V_{i+\HALF}^L =   \omega_0\frac{ V_{i+1} +  V_{i}}{2} 
                  + \omega_1\frac{-V_{i+1} + 3V_{i}}{2} \,,
\end{equation}
where the weights are given by:
\begin{equation}\label{eq:weno3}
  \omega_l = \frac{\alpha_l}{\alpha_0 + \alpha_1}
  \,,\quad
  \alpha_l  = d_l\left(1 + \frac{|\Delta_{i+\HALF}-\Delta_{i-\HALF}|^2}{\beta_l+\epsilon}\right)
  \,,\quad
  \beta_0 = \Delta_{i-\HALF}^2
  \,,\quad
  \beta_1 = \Delta_{i+\HALF}^2\,,
\end{equation}
with $l=\{0,1\}$.
In Eq. (\ref{eq:weno3}), $\Delta_{i+\HALF} = (V_{i+1} - V_i)$ whereas $\{ d_0 = 3/4$, $d_1 =1/4 \}$ (note that in the standard finite volume formulation, starting from 1D cell averages, one has $d_0=2/3$, $d_1=1/3$).
Finally, in order to avoid loss of accuracy at critical points (see \cite{Yamaleev_Carpenter2009} and \cite{Mignone_etal2010}) we adopt here $\epsilon = \Delta x^2$.

\subsection{Flux computation and temporal integration}
\label{sec:flux}
%

Spatial reconstruction provides the $4^{\rm th}$-order accurate pointwise values for left/right states at a given face center (e.g. $V^L_{\xf}$,$V^R_{\xf}$).
These serve as input states to the Riemann solver for the computation of the high-order interface fluxes.
Our implementation comes with a variety of Riemann solvers, such as the linearized Roe solver \citep{Roe1981}, HLL-type Riemann solvers \citep{HLL1983}, including the HLLC \citep{TSS1994, Li2005,Mignone_Bodo2006} and HLLD \citep{Miyoshi_Kusano2005,Mignone_etal2009}, the Lax-Friedrichs \citep{Toro2009}, and the GFORCE \citep{Toro_Titarev2006} solver.
Denoting with $\mathcal{F}_{x,y,z}(\cdot,\,\cdot)$ a generic 1-D Riemann solver along the $\{x,y,z\}$ direction, the interface upwind punctual fluxes are obtained as:
\begin{equation}\label{eq:fluxupwind}
    F_{x,\xf} = \mathcal{F}_x(V^L_\xf, V^R_\xf)  \, , \quad
    F_{y,\xf} = \mathcal{F}_y(V^L_\yf, V^R_\yf)  \, , \quad
    F_{z,\xf} = \mathcal{F}_z(V^L_\zf, V^R_\zf)  \, .
\end{equation}
$4^{\rm th}$-order accurate surface averages are then obtained by the quadrature rule in \refeq{eq:p2v} using the transverse Laplacian operators:
\begin{equation}\label{eq:fluxint}
  \begin{aligned}
    \favx{F} = F_{x,\xf} + \frac{\Delta^{x}_\bot F_{x,\xf}}{24} , \quad
    \favy{F} = F_{y,\xf} + \frac{\Delta^{y}_\bot F_{y,\xf}}{24} , \quad
    \favz{F} = F_{z,\xf} + \frac{\Delta^{z}_\bot F_{z,\xf}}{24}  \, .
  \end{aligned}
\end{equation}
Note that this requires an additional integration in the first layer of ghost zones in order to retrieve the flux information.
Moreover, thanks to the deployment of point value reconstructions, we do not need to solve a Riemann problem with interface-averaged primitive states as it is instead done by Felker \& Stone (Eq. 17 in \citep{Felker_Stone2018}).

These fluxes are used to build the right-hand side of \refeq{eq:ode} which will be hereafter denoted with $\mathcal{L}(U)$. 
Our numerical scheme employs the five stage $4^{\rm th}$-order explicit Strong Stability Preserving Runge-Kutta method (eSSPRK(5,4), \citep{Isherwood_RK42018, Spiteri_Ruuth2002}) to discretize \refeq{eq:ode}
\begin{equation}\label{eq:rk4}
    \begin{array}{lcl}
      U^{(1)} &=& \DS U^n + 0.391752226571890\Delta t \mathcal{L}(U^n)
             \\ \noalign{\medskip}
      U^{(2)} &=& \DS 0.444370493651235U^n + 0.555629506348765U^{(1)} + 0.368410593050371\Delta t \mathcal{L}(U^{(1)})
             \\ \noalign{\medskip}
      U^{(3)} &=& \DS 0.620101851488403U^n + 0.379898148511597U^{(2)} + 0.251891774271694\Delta t \mathcal{L}(U^{(2)})
             \\ \noalign{\medskip}
      U^{(4)} &=& \DS 0.178079954393132U^n + 0.821920045606868U^{(3)} + 0.544974750228521\Delta t \mathcal{L}(U^{(3)})
             \\ \noalign{\medskip}
      U^{n+1} &=& \DS 0.517231671970585U^{(2)} + 0.096059710526147U^{(3)} + 0.063692468666290\Delta t \mathcal{L}(U^{(3)}) 
             \\ \noalign{\medskip}
              & & \DS + 0.386708617503268U^{(4)} + 0.226007483236906\Delta t \mathcal{L}(U^{(4)})\, .
     \end{array}
\end{equation}
The eSSPRK(5,4) method is strongly stable (namely, no bounded temporal growth is allowed: $\Vert U^{n+1} \Vert \leq \Vert U^n \Vert$) under the assumption that the forward Euler method employed is strongly stable under the Courant-Friedrichs-Levy (CFL) restriction ($\Vert U^n + \Delta t \mathcal{L}(U^n) \Vert \leq \Vert U^n \Vert$, \citep{ShuOsher1988, Gottlieb_etal2001}).
In our implementation, the CFL condition determines the time step according to
\begin{equation}
   \Delta t = {\cal C}\DS\min_{ijk}\left[\DS\min_{d}\left(\DS\frac{\Delta x^d}{|\lambda^d_{\rm max}|}\right)\right]
\end{equation}
where ${\cal C}$ is the CFL number, $\lambda^d_{\rm max}$ is the outermost fast magnetotosonic wave and $\Delta x^d$ is the cell length through the different directions $d$.

For a detailed description of the stability properties of the method we refer the reader to \citep{Gottlieb_etal2001, Spiteri_Ruuth2002, Gottliegetal_2009, Gottliebetal_2011, Ketcheson2008}.

\subsection{Order reduction in the presence of discontinuities}

The recovery of point values from volume averages follows the assumption that the function $Q$ is locally continuous up to the $3^{\rm rd}$ derivative in its expansion.
However, this assumption manifestly breaks down in the immediate neighbour of a discontinuity, where using \refeq{eq:v2p} may easily produce unphysical values such as negative energies or densities.
In order to address this shortcoming, we modify the recovery of pointwise conservative variables at the beginning of any integration stage by modifying \refeq{eq:v2p} as
\begin{equation}\label{eq:limited_v2p}
 U_\cc = \DS \av{U}_\cc - \theta_\cc
          \frac{\Delta\av{U}_\cc}{24} + O(h^4) \, ,
\end{equation}
where $\theta_\cc$ has been introduced to detect the presence of a discontinuity
\begin{equation}
    \theta_\cc = \left\{\begin{array}{ll}
      1  & \quad{\rm if}\quad \eta_\cc < \eta_d \,, \\ \noalign{\medskip}
      0  & \quad{\rm otherwise} \,.
    \end{array}\right.
\end{equation}
The parameter $\eta_\cc$ is defined as $\eta_\cc=\sqrt{\eta^2_{\cc,x}+\eta^2_{\cc,y}+\eta^2_{\cc,z}}$, while $\eta_d$ is a threshold parameter.
When $\theta_\cc = 0$ in Eq. \eqref{eq:limited_v2p}, the local order of the scheme is reduced in proximity of discontinuous fronts, including avoiding the integration of fluxes (Eq. \ref{eq:fluxint}).
In these regions, we further increase the robustness of the scheme by also lowering the order of the reconstruction scheme by switching to the $3^{\rm rd}$-order WENO scheme (see end of \S\ref{sec:point_rec}) or to piecewise linear reconstruction.
This approach (known as the \quotes{fallback}) has been employed, in a similar fashion, also by \cite{Nunez_2016a} in the context of high-order schemes.
We now propose two different discontinuity detectors in order to evaluate $\eta_\cc$.

The first one is based on Jameson's pressure-based shock sensor \cite{Jameson_1981}
\begin{equation}\label{eq:ho_jameson_lim}
  \eta_{\cc,x} = \frac{\av{Q}_{\cc+\hvec{e}_x} - 2\av{Q}_{\cc} + \av{Q}_{\cc-\hvec{e}_x}} 
                {|\av{Q}_{\cc+\hvec{e}_x}| + 2|\av{Q}_{\cc}| + |\av{Q}_{\cc-\hvec{e}_x}| +  \epsilon}    \,,
\end{equation}
(similar expressions hold for $\eta_{\cc,y}$ and $\eta_{\cc,z}$) where $\epsilon$ is a small number to prevent negligible variations from triggering the switch.
\refeq{eq:ho_jameson_lim} is first evaluated with $\av{Q}$ being the volume-averaged density, pressure, and magnetic pressure and the maximum over these three values is finally considered.
Jameson's switch has the advantage of keeping the stencil compact (no additional ghost zones are needed), albeit three zones may not be sufficient to distinguish a discontinuity from a smooth extremum \citep{Suresh_Huynh1997}.

The second discontinuity sensor that we propose employs a derivatives ratio that attempts to identify high-frequency oscillations which typically arise near a discontinuous front. 
Since high wave-number Fourier modes will not be dumped around these critical points, we expect the contribution of higher order derivatives to be non-negligible. 
To this end, we first evaluate the normalized contribution of odd and even modes separately by evaluating the parameters $\eta^{o}_\cc$ and $\eta^{e}_\cc$
\begin{equation}\label{eq:ho_rder_lim}
  \eta^{o}_\cc = \frac{|\delta^{(3)} \av{Q}_\cc|}
                  {|\av{Q}_{\cc, {\rm ref}}| + |\delta^{(1)} \av{Q}_\cc| 
                                             + |\delta^{(3)} \av{Q}_\cc| + \epsilon}
  \,,\qquad
  \eta^{e}_\cc = \frac{|\delta^{(4)}\av{Q}_\cc|}
                  {|\av{Q}_{\cc, {\rm ref}}| + |\delta^{(2)} \av{Q}_\cc| 
                                             + |\delta^{(4)} \av{Q}_\cc| + \epsilon} \, ,
\end{equation}
where $\delta^{(m)}Q_\cc$ (with $m=1,..,4$) are undivided $2^{\rm nd}$-order accurate approximations to the $m$-th derivative.
In the $x$-direction, for instance:
\begin{equation}\begin{array}{lcl}
  \delta^{(1)} \av{Q}_\cc &=& 
               \DS \frac{1}{2}\left(\av{Q}_{\cc+\hvec{e}_x} - \av{Q}_{\cc-\hvec{e}_x}\right) \, , \\ \noalign{\medskip}
  \delta^{(2)} \av{Q}_\cc &=& 
               \DS{\av{Q}_{\cc+\hvec{e}_x} - 2\av{Q}_\cc + \av{Q}_{\cc-\hvec{e}_x}} \, , \\ \noalign{\medskip}
  \delta^{(3)} \av{Q}_\cc &=& 
               \DS \frac{1}{2} \left(\av{Q}_{i+2\hvec{e}_x} - 2(\av{Q}_{\cc+\hvec{e}_x} - 2\av{Q}_{\cc-\hvec{e}_x}) -\av{Q}_{\cc-2\hvec{e}_x}\right)  \, ,\\ \noalign{\medskip}
  \delta^{(4)} \av{Q}_\cc &=& 
               \DS \av{Q}_{\cc+2\hvec{e}_x} - 4\av{Q}_{\cc+\hvec{e}_x} + 6\av{Q}_\cc - 4 \av{Q}_{\cc-\hvec{e}_x} + \av{Q}_{\cc-2\hvec{e}_x} \, .
  \end{array}
\end{equation}
In our experience, using primitive variables has shown to yield a more robust identification method. 
For this reason, we take $\av{Q}_{\cc}={\cal V}(\av{U}_\cc)$ at the beginning of the Runge-Kutta stage.
The reference value $\av{Q}_{\cc, \rm ref}$ in Eq. (\ref{eq:ho_rder_lim}) is equal to $\av{Q}_{\cc}$ for density and pressure while we employ $\sqrt{p_\cc/\rho_\cc}$ (isothermal sound speed) for velocity components and $|B|_\cc$ for magnetic field components.
Once $\eta^o_\cc$ and $\eta^e_\cc$ have been computed, we simply take the maximum of the two: $\eta_\cc = \max (\eta_\cc^o,\eta_\cc^e)$.

The derivative ratio detector needs a larger stencil thereby asking for one additional ghost zone.
Along the same arguments, a more sophisticated approach that requires an even larger stencil, is shown by Bambozzi \& Pires \cite{Bambozzi_Pires_2016}.
\section{The $4^{\rm th}$-order constrained transport}
\label{sec:ct}
%

\begin{figure}[!ht]
  \centering
  \includegraphics[trim={90 10 50 10}, width=0.5\textwidth]{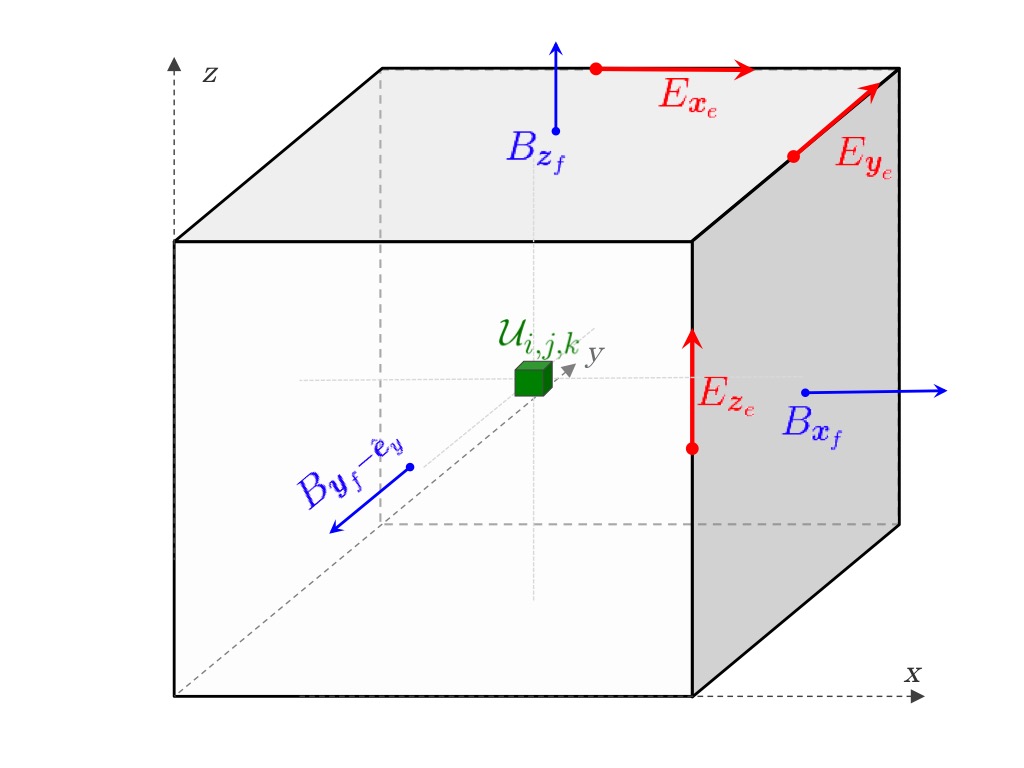}%
  \includegraphics[width=0.55\textwidth]{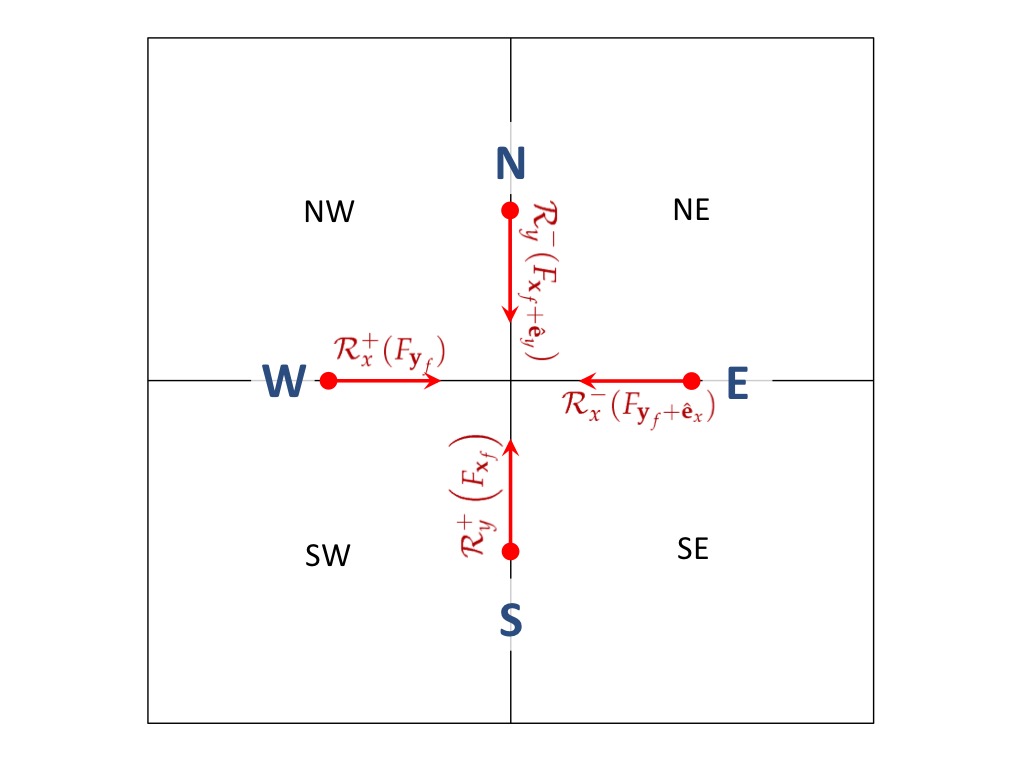}
  \caption{\footnotesize Left: positioning of MHD variables in the UCT formalism.
   Staggered magnetic field components (blue) are face-centered, the electromotive force is edge-centered  (red) while remaining HD quantities are located at the zone center (green).
   Right: top view of the intersection between four neighbor zones: N, S, E, and W indicate the four cardinal directions with respect to the zone edge (here represented by the intersection between four neighbor zones), $\cR_x(F_{\yf})$ and $\cR_y(F_{\xf})$ are 1-D reconstruction operators applied to each zone face, see \refeq{eq:edge_flux}.}
  \label{fig:ct}
\end{figure} 

As we outlined in \S \ref{sec:equations}, the HD sub-system cannot be straightforwardly extended to the MHD system without introducing numerical strategies to control the rising of spurious magnetic monopoles.
Failure to fulfill this requisite may lead to unphysical effects such as plasma acceleration parallel to the field, incorrect jump conditions, wrong propagation speed of discontinuities and odd-even decoupling \citep{Brackbill_Barnes1980, Toth2000,  Balsara_Spicer1999}.
In what follows we employ the upwind constrained transport method \citep{Evans_Hawley1988, Dai_Woodward1998, Balsara_Spicer1999, Ryu_etal1998, Londrillo_DelZanna2004, Mignone_DelZanna2021} in order to ensure that the divergence-free condition is maintained to machine accuracy.

In the FV-CT formalism the field's surface integrals $\favx{B}$, $\favy{B}$, $\favz{B}$ are the primary representation of the magnetic field evolved in time.
\refeq{eq:induction} regulates the time variation of the magnetic field fluxes, implying that the formal correspondence
\begin{equation}\label{eq:edge_flux}
  \begin{array}{lll}
    F^{[B_x]}_x =  0     \,,\;  &
    F^{[B_x]}_y =  E_z   \,,\;  &
    F^{[B_x]}_z = -E_y   \,,
  \\ \noalign{\medskip}
    F^{[B_y]}_x = -E_z  \,,\;  &
    F^{[B_y]}_y =    0  \,,\;  &
    F^{[B_y]}_z =  E_x  \,,
  \\ \noalign{\medskip}
    F^{[B_z]}_x =  E_y   \,,\;  &
    F^{[B_z]}_y = -E_x   \,,\;  &
    F^{[B_z]}_z =    0   \,,
  \end{array}
\end{equation}
holds when solving a Riemann problem.
While this certainly holds for the fluxes computed at the face midpoint, \refeq{eq:stokes} requires the EMF to become available at a zone edge to serve as a discrete version of Stokes theorem.
In order to reconstruct the induction fluxes from the face center (where they are available as point values) to one of the four adjacent edges,  we employ the UCT method of Mignone \& Del Zanna \citep{Mignone_DelZanna2021}.
This method provides a general formalism for systematic construction of EMF averaging procedures where dissipation terms depend on the left and right transverse magnetic field components.

Referring to the right panel of \refig{fig:ct} and being consistent with the notation presented in Mignone \& Del Zanna \citep{Mignone_DelZanna2021}, we consider a top view of the intersection of four zones at a cell-edge.
We denote, respectively, the left and right reconstructed states along the y-coordinate as south (S) and north (N), while on the x-direction the reconstructed states are labeled with west (W) or east (E) with respect to the intersection point.
The edge-centered EMF with the desired upwind properties can be build up from
\begin{equation}\label{eq:emf}
    E_{z,\ze} = - \left[\left(a_x \overline{v}_x B_y\right)^W 
                    + \left(a_x \overline{v}_x B_y\right)^E\right]
              + \left[\left(a_y \overline{v}_y B_x\right)^N 
                    + \left(a_y \overline{v}_y B_x\right)^S\right]
              + \left[\left(d_x B_y \right)^E - \left(d_x B_y \right)^W\right]
              - \left[\left(d_y B_x \right)^N - \left(d_y B_x \right)^S\right] \, ,
\end{equation}
where $\overline{v}_{x,y}$ are the transverse velocities reconstructed from the interfaces, $a_{x,y}$ and $d_{x,y}$ are the flux and diffusion coefficients, and $B_{x,y}$ are the pointwise values of the magnetic field reconstructed from the corresponding interface.
The other EMF components may be obtained by cyclic permutations. 
For the sake of briefness, we address the reader to find all the details of the different EMF averaging techniques that we employ in the present work in Mignone \& Del Zanna \citep{Mignone_DelZanna2021}.
Note that the employment of point value reconstruction methods (rather than 1D volume average-based ones) requires the electric field to be evaluated over 3 layers of ghost zones in the transverse direction rather than 4.

Finally, the line-averaged integrated electric field is computed as
\begin{equation} \label{eq:emfint}
  \lavx{E} = \DS E_{x,\xe} + \frac{\Delta^{x} E_{x,\xe}}{24} , \quad
  \lavy{E} = \DS E_{y,\ye} + \frac{\Delta^{y} E_{y,\ye}}{24} , \quad
  \lavz{E} = \DS E_{z,\ze} + \frac{\Delta^{z} E_{z,\ze}}{24} \, ,
\end{equation}
and then employed to update the staggered magnetic field components using the discrete version of Stoke's theorem \refeq{eq:stokes}.

In the following we recap the steps of our $4^{\rm th}$-order numerical scheme.
\begin{enumerate}
    \item Start with the volume average of conserved quantities $\av{U}_{\cc}$ and the face average of the staggered magnetic field components, $\favx{B}^x$, $\favy{B}^y$, $\favz{B}^z$.
    \item Assign boundary conditions to these variables.
    \item Obtain volume-averaged primitive variables by converting the volume averages $\av{V}_c = {\cal V}(\av{U}_\cc)$. 
    Notice that this array is $2^{\rm nd}$-order accurate and it is only used in the limiting process, thus it does not replace the pointwise conversion later on.
    \item Evaluate the $\theta_\cc$ parameter for order reduction in presence of discontinuities by means of either the Jameson's shock sensor (Eq. \ref{eq:ho_jameson_lim}) or by weighing up the contribution of higher-order derivatives (Eq. \ref{eq:ho_rder_lim}) and flag troubled cells along with first neighbours.
    \item Get pointwise value $U_\cc$ (Eq. \ref{eq:limited_v2p}) at the cell center and  $B_{\xf}, B_{\yf}, B_{\zf}$ (Eq. \ref{eq:v2p_B}) at the corresponding face centers.
    Also, during this step, interpolate the magnetic field from the face center to the cell center (Eq. \ref{eq:Bcp}).
    \item Convert pointwise conservative variables $U_\cc$ in primitive variables $V_\cc = {\cal V}(U_\cc)$.
    \item Reconstruct L/R interface values (Eq. \ref{eq:mp5} or \ref{eq:wenoz}) in primitive (or characteristic) variables.
    \item Obtain upwind fluxes of hydro quantities (Eq. \ref{eq:fluxupwind}) and average them over the corresponding faces (Eq. \ref{eq:fluxint}).
    Concurrently, store the transverse velocities (e.g., $\overline{v}_y$ and $\overline{v}_z$ at an $x$-interface) as well as the coefficients $a^{L/R}$ and $d^{L/R}$ available with the chosen Riemann solver at a zone interface.
    These quantities are needed in the next step of the UCT scheme. For completeness we report the explicit form of these coefficients in \ref{app:uct_averages}.

    \item Reconstruct the transverse velocities obtained with the 1D Riemann solver and the normal components of the magnetic field from the face centers to the edges.
    This will yield the $N,S$ and $W,E$ values in Eq. (\ref{eq:emf}).
    At an $x$-interface, for instance, one has to reconstruct $\overline{v}_{y,\xf}$ and $B_{x,\xf}$ along the $y$-direction, and then, again, $\overline{v}_{z,\xf}$ and $B_{x,\xf}$ along the $z$-direction.
    
    \item Obtain the electric field at corners (Eq. \ref{eq:emf}) by means of different averaging procedures (see \ref{app:uct_averages}).
    \item Line-average the electric field according to Eq. (\ref{eq:emfint}).
    \item Add all contributions to the right-hand side and obtain the solution $\av{U}_{\cc}$ at the next time level or RK stage (Eq. \ref{eq:rk4}).
\end{enumerate}
\section{Numerical benchmarks}
\label{sec:numerical_benchmarks}
%
%
%

We now test the capabilities of the novel method in a series of 2D and 3D numerical benchmarks.
Each test is validated twice, by solving for the MHD and the RMHD case.
As a reference, we will compare our results against the traditional $2^{\rm nd}$- or $3^{\rm rd}$-order schemes.
As a default, \refeq{eq:ode} will be discretized by the $2^{\rm nd}$-, $3^{\rm rd}$-, or $4^{\rm th}$-order schemes by means of a RK2, RK3, or RK4 time integrator.
The WENOZ and MP5 spatial reconstructions will always use point values and, otherwise differently stated, each MHD and RMHD benchmark will employ, respectively, a HLLD solver coupled with the UCT-HLLD EMF average, and the GFORCE solver plus the UCT-GFORCE average that we suitably extended to relativistic frameworks.

Whenever the initial condition is a smooth analytic function our scheme integrates it by means of a Gaussian quadrature rule employing 4 Gaussian points.

We evaluate $L_1$ norm errors for a generic quantity $Q$ against a reference solution $Q_{\rm ref}$ as $\epsilon_1(Q) = \sum_i|Q(x_i)-Q_{\rm ref}(x_i)|/N$, where $x_i$ is a generic point of the domain and $N$ is the total number of grid zones.
The convergence rate $O_{L_1}$ is thus calculated via a linear fit of $\log(\epsilon_1)$ against $\log(N_x)$, with $N_x$ being the linear resolution.

All computations have been carried out by means of the PLUTO astrophysical code \citep{Mignone_PLUTO2007, Mignone_PLUTO2012}, where the $4^{\rm th}$-order method has been implemented.

\subsection{3D Rotated Shock Tube} \label{sec:sct}

\begin{figure}[!h] 
\centering
    \includegraphics[width=0.90\textwidth]{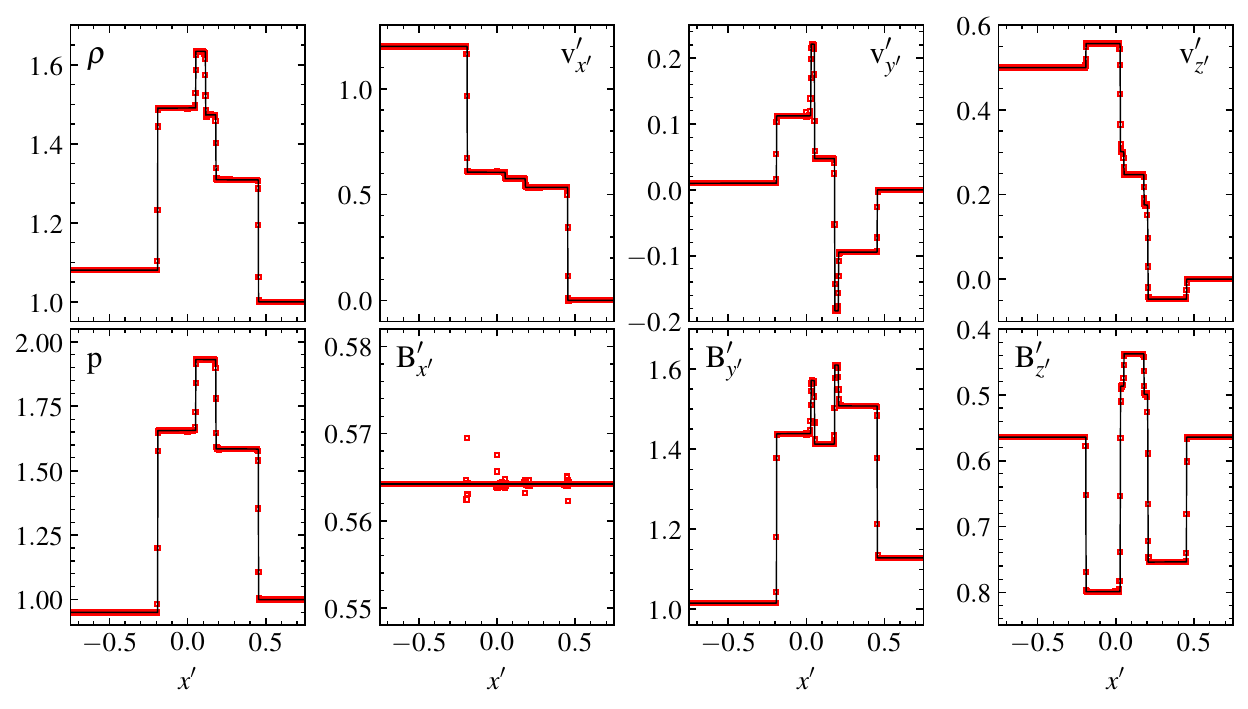}%
    \caption{\footnotesize Primitive variable profiles for the 3D rotated MHD shock tube problem at $t_f = t'_f\cos\alpha\cos\gamma$, where $t'_f = 0.2$ obtained with the WENOZ $4^{\rm th}$-order scheme (red squares) on a uniform grid of $768\times8\times8$ zones.
    All the variables are plotted along the non-rotated direction $x'$.
    On the left panels density (top) and pressure (bottom) are shown, while in the remaining panels, the three components of velocity (top) and magnetic field (bottom) are plotted.
    The black solid lines represent a reference 1D solution computed by emplyoing a $2^{\rm nd}$-order scheme with a resolution of $N_x = 16384$ grid points.}
    \label{fig:st_MHD}
\end{figure}

\begin{figure}[!h] 
\centering
    \includegraphics[width=0.90\textwidth]{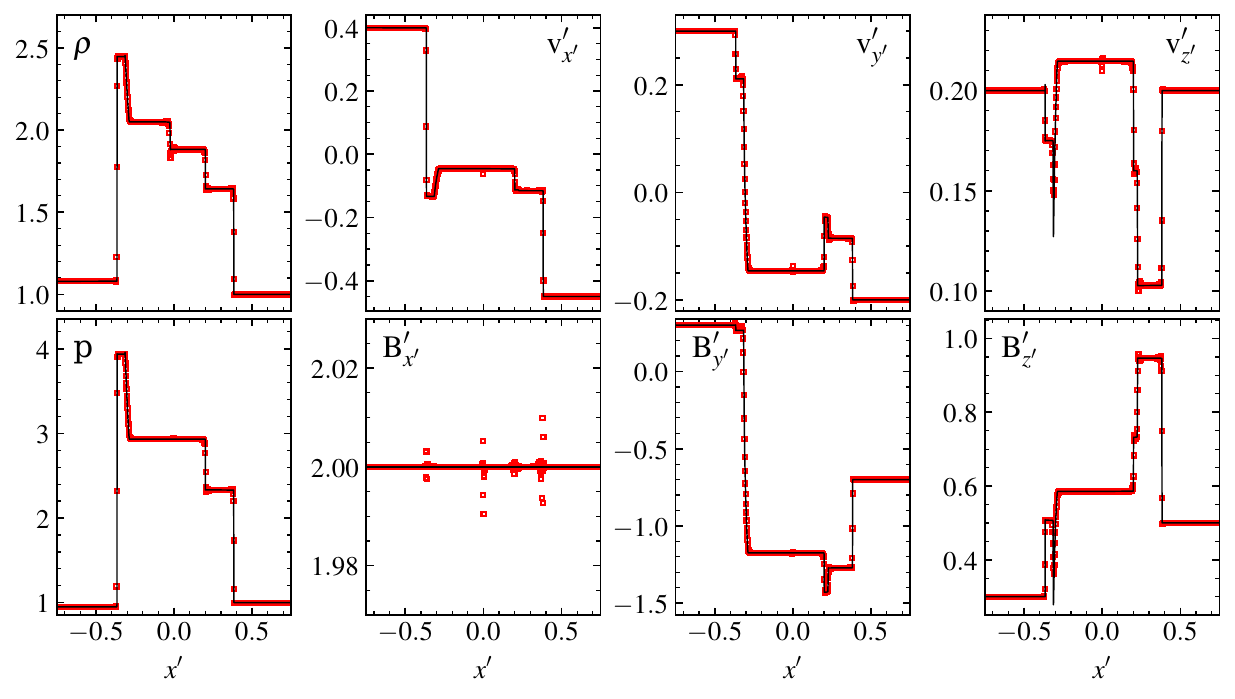}%
    \caption{\footnotesize Same of Fig. \ref{fig:st_MHD} but for the relativistic case.
    Here the numerical scheme adopted is using the MP5 algorithm and the ``unrotated'' final time is $t_f' = 0.55$.}
    \label{fig:st_RMHD}
\end{figure}

We begin this section by considering a 3D rotated version of a standard shock-tube problem.
The initial condition is best expressed in the original (non-rotated) frame where $x'$ is the coordinate perpendicular to the plane of discontinuity.
In the MHD case we have
\begin{equation}\label{eq:mhd_rotated_sod}
  \left\{\begin{array}{lclr} 
    \cV'_L &=&\DS \Big(1.08,\,1.2,\,0.01,\,0.5,\, 
                       \frac{2}{\sqrt{4\pi}},\,  
                       \frac{3.6}{\sqrt{4\pi}},\, 
                       \frac{2}{\sqrt{4\pi}},\, 0.95 \Big)
          & \quad{\rm for}\quad x' < 0 \, ,
    \\ \noalign{\medskip}
    \cV'_R &=&\DS \Big(1,   \,  0,\,   0,\,  0,\, 
                       \frac{2}{\sqrt{4\pi}},\, 
                       \frac{4}{\sqrt{4\pi}},\,   
                       \frac{2}{\sqrt{4\pi}},\, 1\Big)
          & \quad{\rm for}\quad x' > 0 \, ,
  \end{array}\right.
\end{equation}
where $\cV' = (\rho, v'_{x'}, v'_{y'}, v'_{z'}, B'_{x'}, B'_{y'}, B'_{z'}, p)$ is an array of primitive quantities in the non-rotated frame.
In the relativistic case we have, likewise,
\begin{equation}\label{eq:rmhd_rotated_sod}
  \left\{\begin{array}{lclr} 
    \cV'_L &=&\DS \left(1.08,\,0.4,\,0.3,\,0.2,\, 
                        2,\,  0.3,\, 0.3,\, 0.95 \right)
          & \quad{\rm for}\quad x' < 0\, ,
    \\ \noalign{\medskip}
    \cV'_R &=&\DS \left(1,   \,-0.45,\, -0.2,\,  0.2,\, 2,\, -0.7,\, 0.5,\, 1\right)
          & {\rm for}\quad x' > 0\, .
  \end{array}\right.
\end{equation}
For each case, we rotate the initial condition by an angle $\alpha$ around the $z$-axis and then by an angle $\gamma$ around the $y$-axis.
The coordinate transformation used for the 3D rotation is explained in detail in \cite{Mignone_etal2010} (see \S 4.2.2 in that paper) and leads to a planar symmetry condition whereby, for any quantity $q(\vec{x})$, one must have $q(\vec{x}+\vec{s}) = q(\vec{x})$, where $\vec{s} = (n_x\Delta x,\, n_y\Delta y,\, n_z\Delta z)$ corresponds to an integer shift of cells.
We choose $\tan\alpha=1/2$ and $\tan\gamma=\cos\alpha\tan\beta = -1/(2\sqrt{5})$ with $\tan\beta = -1/4$ corresponding to a translational invariance by an integer shift of cells, i.e., $(n_x,\, n_y,\,n_z) = (1,-2,0)$ and $(n_x,\, n_y,\,n_z) = (1,0,4)$.

We solve the MHD / RMHD equations using our $4^{\rm th}$-order method, adopting the WENOZ reconstruction in characteristic variables for the classical version and  the MP5 reconstruction in primitive variables for the relativistic variant.
The computational domain is the box $[-L/2,\, L/2] \times [0,\,L/96] \times [0,\,L/96]$ covered by a uniform grid of $768\times8\times8$ zones.
We evolve the solution until $t_f = t'_{f}\cos\alpha\cos\gamma = 4/\sqrt{21}$ where $t'_f=0.2$ (for classical MHD), $t'_f = 0.55$ is the final time for the non-rotated (1D) versions of the problem, see, e.g., \cite{Ryu_etal1995, Gardiner_Stone2008, Mignone_etal2010, Mignone_etal2010} for the classical MHD version and \cite{Balsara_2001, Mignone_etal2009, Anton_2010} (and references therein) for the relativistic variant. 
One dimensional plots of relevant fluid quantities are shown in the unrotated frame in Fig. (\ref{fig:st_MHD}) and (\ref{fig:st_RMHD}) against a 1D high-resolution reference profile computed with a $2^{\rm nd}$-order scheme with a resolution of $N_x = 16384$ grid points. 
The solution features a contact discontinuity in the middle ($x\approx 0$) separating two outermost fast magnetosonic shocks enclosing a pair of rotational waves followed by two slow waves. 
Profiles are correctly reproduced and oscillations in the normal component of magnetic field are of the same order as those of \cite{Mignone_etal2010} and smaller than those in \cite{Gardiner_Stone2008} who used a $2^{\rm nd}$-order scheme.
Our results demonstrate that our $4^{\rm th}$-order method can correctly capture propagating discontinuous fronts although the order of the scheme degenerates to $1^{\rm st}$-order.

\subsection{2D iso-density and iso-enthalpy vortex} \label{sec:iso}

\begin{figure}[!h] 
\centering
    \includegraphics[width=0.90\textwidth]{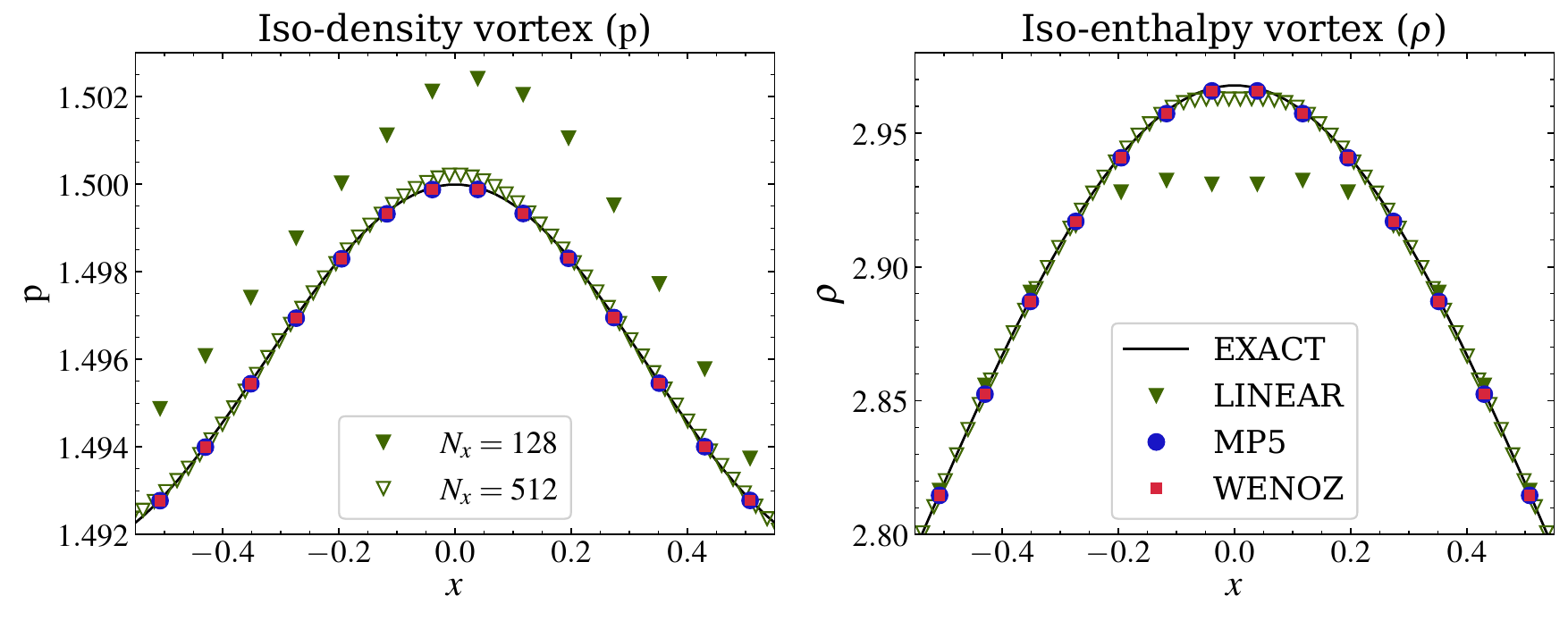}%
    \caption{\footnotesize Zoomed plot of the thermal pressure profile in the 2D MHD iso-density vortex (left panel) and density in the 2D RMHD iso-enthalpy vortex (right panel).
    In each panel the exact solution's profile (continuous black line) is compared with the profiles evolved by a $4^{\rm th}$-order scheme adopting the WENOZ (red squares), the MP5 (blue circles), and the $2^{\rm nd}$-order LINEAR scheme (green triangles).
    Filled symbols refer to data produced at a resolution of $N_x = 128$, while the empty ones are data at $N_x = 512$.}
    \label{fig:iso-vortex}
\end{figure}

We consider the MHD iso-density vortex advection problem (as introduced by Balsara \citep{Balsara2004} and later presented by Mignone \citep{Mignone_etal2010} and Dumbser \citep{Dumbser_2008}) and also propose a new exact equilibrium solution of the ideal RMHD equations with constant enthalpy.
For the MHD case, the initial condition consists of a magnetized vortex at constant density in force balance that propagates along the main diagonal of a computational domain of $[-5, 5] \times [-5, 5]$ under periodic boundary conditions.
The equilibrium condition is best expressed in cylindrical coordinates $(R, \phi, z)$, from radial momentum balance
\begin{equation}\label{eq:vortex_rad_equilMHD}
    \DS \frac{d}{dR} \left(p + \frac{B_z^2}{2} \right) = 
        \frac{\rho v^2_{\phi}}{R} - \frac{1}{2R^2} \frac{d}{dR}\left(R^2B^2_{\phi}\right) \, .
\end{equation}
We assume constant density $\rho = 1$ and \refeq{eq:vortex_rad_equilMHD} is satisfied by adopting the following profiles:
\begin{equation}\label{eq:vortex_v_and_B}
   \vec{v} = \vec{v}_0 + (-y, x)ke^{q(1-R^2)}
   \,,\quad
   \vec{B} = (-y, x)\mu e^{q(1-R^2)} + B_z\hvec{e}_z\, ,
\end{equation}
where $\vec{v}_0$ is the translational velocity, $R = \sqrt{x^2 + y^2}$ is the cylindrical radius, $k = \mu = 1/2\pi$ and the parameter $q$ is set to $1$ as in \citep{Dumbser_2008, Mignone_etal2010}.
The pressure term on the left hand side of Eq. (\ref{eq:vortex_rad_equilMHD}) is then readily obtained from 
\begin{equation} \label{eq:vortex_pt}
  p_{\rm t}\equiv p + \frac{B_z^2}{2} = 1 + \frac{1}{4q}
\left[\mu^2\left(1-2qR^2\right) - f k^2\right]e^{2q(1-R^2)} \,,
\end{equation}
where $f = \rho$ for the classical case.
We equally distribute $p_t$ in Eq. (\ref{eq:vortex_pt})  among the thermal and magnetic contributions, i.e. $p = \alpha p_{\rm t}$ and $B_z^2/2 = (1-\alpha)p_{\rm t}$, where $\alpha \in [0, 1]$.
In this setup, we use $\alpha=1/2$ and $\vec{v}_0 = (1, 1)$.

In the RMHD case (for which we also assume a purely radial dependence), the time-independent radial component of the momentum equation takes instead the form
\begin{equation}\label{eq:vortex_rad_equilRMHD}
    \DS \frac{dp}{dR} = \frac{w\gamma v_{\phi}^2}{R} + \frac{E_R}{R}\frac{d\left(R E_R\right)}{dR} - B_z\frac{dB_z}{dR} - \frac{B_{\phi}}{R}\frac{d\left(RB_{\phi}\right)}{dR} \,,
\end{equation}
where $w=5$ is the constant enthalpy. 
We assume $B_z = E_R = 0$, so that Eq. (\ref{eq:vortex_rad_equilRMHD}) has solution given by Eq. (\ref{eq:vortex_v_and_B}) and (\ref{eq:vortex_pt}) with  $f=w$.
Here we choose a static vortex ($v_0=0$).
Density is recovered from the ideal equation of state (adiabatic index $\Gamma=5/3$) as $\rho=\DS w - 5p/2$, which is ensured to be strictly positive by our choice of the initial enthalpy.

\begingroup
\begin{table}
\centering
\begin{tabular}{cccccccccccc}
\hline
  $N_x$ &  \multicolumn{2}{c}{MHD MP5} & & \multicolumn{2}{c}{MHD WENOZ}  & &  \multicolumn{2}{c}{RMHD MP5} & & \multicolumn{2}{c}{RMHD WENOZ}  \\ 
\cline{2-3} \cline{5-6} \cline{8-9} \cline{11-12}
          & $\epsilon_1(B_x)$  & $O_{L_1}$ & & $\epsilon_1(B_x)$  & $O_{L_1}$ 
        & & $\epsilon_1(B_x)$  & $O_{L_1}$ & & $\epsilon_1(B_x)$  & $O_{L_1}$\\    
\hline
\hline
$32$   &  9.99E-04  &   -    & &  9.01E-04  &   -     & 
       &  3.93E-04  &   -    & &  3.11E-04  &   -     \\
 $64$  &  6.09E-05  &  4.04  & &  6.00E-05  &  3.91    & 
       &  1.70E-05  &  4.53  & &  1.83E-05  &  4.09   \\        
 $128$ &  2.35E-06  &  4.70  & &  2.37E-06  &  4.66    & 
       &  6.09E-07  &  4.80  & &  6.41E-07  &  4.84   \\
 $256$ &  1.06E-07  &  4.47  & &  1.06E-07  &  4.48    & 
       &  2.34E-08  &  4.71  & &  2.36E-08  &  4.77   \\
 $512$ &  5.76E-09  &  4.20  & &  5.76E-09  &  4.20    & 
       &  1.05E-09  &  4.47  & &  1.05E-09  &  4.49   \\
\hline
\end{tabular}
\caption {\footnotesize $L_1$ norm errors and corresponding convergence rates of the 2D MHD iso-density vortex (columns 2-5) and the 2D RMHD iso-enthalpy vortex (columns 6-9).
Computations were carried out with the resolution of 32, 64, 128, 256 and 512 grid points along each direction.}\label{tab:iso_vortex}
\end{table}
\endgroup
Both the classical and relativistic cases are evolved until $t=10$, which corresponds to one period for the MHD case.
Table~\ref{tab:iso_vortex} reports the errors for $B_x$ measured in $L_1$ norm and the corresponding convergence rates both for the MHD and RMHD solutions by either using the WENOZ or the MP5 reconstructions. 

\refig{fig:iso-vortex} overlays the exact pressure profile (continuous black line) on the data produced with the $4^{\rm th}$-order scheme adopting the WENOZ (red squares), the MP5 (blue circles) and the $2^{\rm nd}$-order LINEAR scheme (green triangles). 
Filled symbols refer to data produced at a resolution of $N_x = 128$, while empty ones are produced at $N_x = 512$ and have been presented only for the LINEAR scheme, since high-order data do not display a sensible difference between the resolution of $128$ and $512$.
The profiles of the $4^{\rm th}$-order methods overlap almost perfectly on the exact solution at $128$, while the $2^{\rm nd}$-order data do not reach such accuracy neither at $128$ nor at $512$ grid points.
Especially in the relativistic case, it is relevant to notice the significant amount of clipping introduced by the LINEAR scheme.
On the other hand, the $4^{\rm th}$-order method is capable of attaining significantly improved accuracy already at modest resolutions.

A close comparison with the results of the finite difference (FD) scheme of Mignone \cite{Mignone_etal2010} reveals very similar errors for $N_x \lesssim 128$ even if the FD method, being $5^{\rm th}$-order accurate, converges faster than the proposed FV method.
We also point out that, owing to the employment of a $3^{\rm rd}$-order Runge Kutta method, the time step of the FD scheme has to be re-scaled as $\Delta t \sim \Delta x^{5/3}$ in order to retain $5^{\rm th}$-order accuracy (see Eq. [48] of that paper).


\subsection{Circularly polarized Alfv\'en waves} \label{sec:cp}

We now consider the oblique propagation of circularly polarized Alfv\'en waves, which is an exact non-linear solution of the (R)MHD equations and was first presented by T\'oth \cite{Toth2000} in the context of classical MHD 
and later by Del Zanna \cite{DelZanna_etal2007} for the relativistic case.

In the MHD case we draw upon the setup of \cite{Mignone_etal2010} by first constructing a 1D wave profile in the $x$-direction with angular frequency $\omega$, wavenumber $k$ and amplitude $\eta$ as
\begin{equation}
    \left(\begin{array}{c}
        v_x  \\ 
        v_y  \\
        v_z    
    \end{array}\right) = 
    \left(\begin{array}{c}
        0  \\ 
        \eta\sin{\phi}  \\
        \eta\cos{\phi}    
    \end{array}\right) \,,\quad
    \left(\begin{array}{c}
        B_x  \\ 
        B_y  \\
        B_z    
    \end{array}\right) = 
    \left(\begin{array}{c}
        c_a\sqrt{\rho} \\ 
        \sqrt{\rho}\eta\sin{\phi} \\
        \sqrt{\rho}\eta\cos{\phi}    
    \end{array}\right) \,,\quad
\end{equation}
where $\phi=kx-\omega t$, $c_a=\omega/k=1$ is the phase (Alfv\'en) velocity and $\eta = 0.1$ is the wave amplitude.
Density and pressure are set to $\rho=1$ and $p=2$, respectively, while we employ the ideal EoS with $\Gamma=5/3$.
Such solution is then rotated by 45$^\circ$ around the $y$-axis in the 2D setup and by an additional 45$^\circ$ around the $z$-axis in the 3D case to have the wave propagating along the computational box's diagonal.
The reader should refer to \cite{Mignone_etal2010} for the details.

\begin{figure}[!ht]
    \centering
    \includegraphics[width=0.95\textwidth]{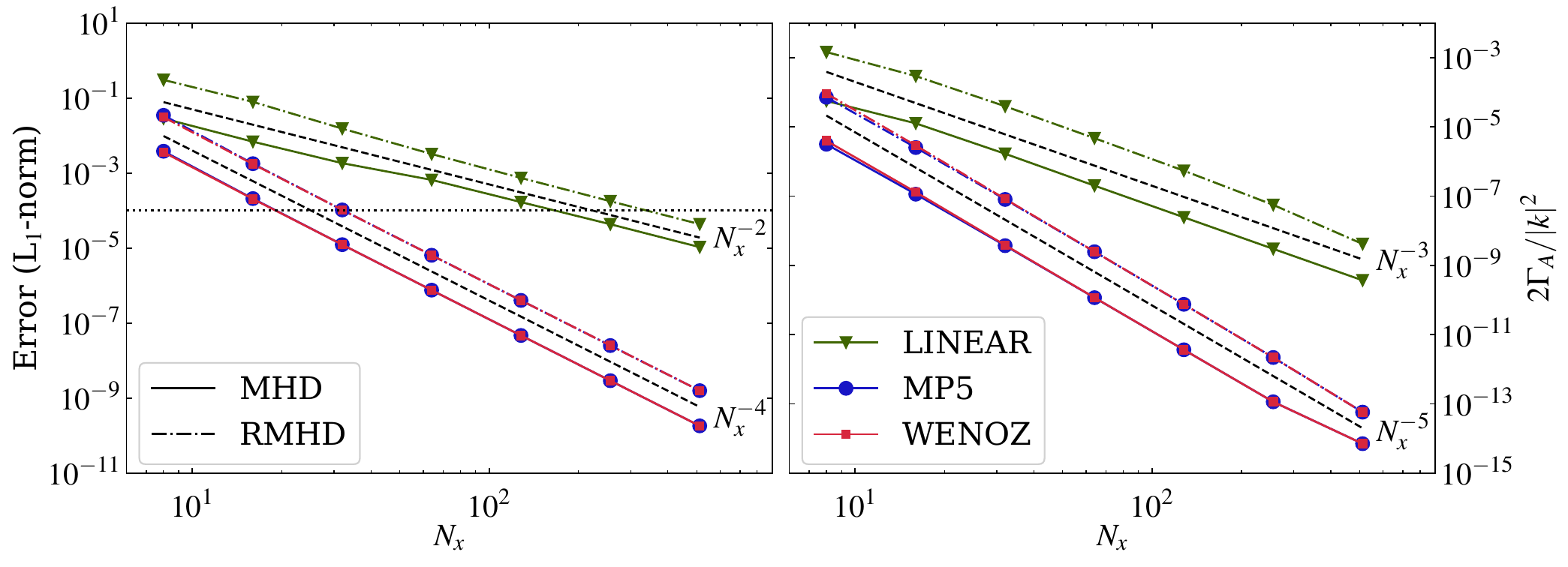}
    \caption{$L_1$-norm errors (left panel) and numerical diffusion (right panel) for the 3D circularly polarized Alfv\'en wave test in the MHD regime (solid lines) and RMHD (dashed-dotted lines) as functions of the grid resolution $N_x$ with $2^{\rm nd}$- and $4^{\rm th}$-order numerical schemes.
    The horizontal dotted line in the left panel is positioned at $10^{-4}$ in order to highlight the different resolution that achieves an accuracy of $10^{-4}$ with either a low or a high order scheme.
    The dashed lines give the ideal convergence slope (left panel) and a reference slope for the numerical diffusion (right panel).
    }\label{fig:cpa_scaling}
\end{figure}

The RMHD case is set in a similar way, with a 1D wave profile defined by (see \cite{DelZanna_etal2007}) 
\begin{equation}
    \left(\begin{array}{c}
        v_x  \\ 
        v_y  \\
        v_z    
    \end{array}\right) = 
    \left(\begin{array}{c}
        0  \\ 
        - c_A \eta\cos{\phi}  \\
        - c_A \eta\sin{\phi}    
    \end{array}\right) \,,\quad
    \left(\begin{array}{c}
        B_x  \\ 
        B_y  \\
        B_z    
    \end{array}\right) = 
    \left(\begin{array}{c}
        B_0 \\ 
        B_0 \eta\cos{\phi} \\
        B_0 \eta\sin{\phi}    
    \end{array}\right) \,,\quad
\end{equation}
where we set $B_0=1$, $\eta=1$, $\Gamma=4/3$ and the Alfv\'en velocity is defined by
\begin{equation}
    c_A = \frac{B_0^2}{\rho h+B_0^2(1+\eta^2)}\left[\frac{1}{2}
          \left( 1+\sqrt{1-\left( \frac{2\eta B_0^2}{\rho h+B_0^2(1+\eta^2)}\right)^2 }\right) \right]^{-1} \,.
\end{equation}
We then apply the same procedure as for the MHD case \citep{Mignone_etal2010} to rotate the wave front.

The computational domain is the unit square (cube) in 2D (3D), while we follow the evolution of the wave for one period, after which its profile is supposed to return exactly on the initial values.
We then estimate the error introduced by the numerical algorithms using the $L_1$ norm evaluated for the $y$- component of the magnetic field between the final and initial profiles.

\begingroup
\begin{table}[!b]
\centering
\begin{tabular}{ccccccccccccc}
\hline
 Scheme & $N_x$ &  \multicolumn{2}{c}{2D MP5} & & \multicolumn{2}{c}{2D WENOZ}  & &  \multicolumn{2}{c}{3D MP5} & & \multicolumn{2}{c}{3D WENOZ}  \\ 
\cline{3-4} \cline{6-7} \cline{9-10} \cline{12-13}
        & &  $\epsilon_1(B_y)$  & $O_{L_1}$ & &  $\epsilon_1(B_y)$  & $O_{L_1}$ 
        & & $\epsilon_1(B_y)$  & $O_{L_1}$ & & $\epsilon_1(B_y)$  & $O_{L_1}$ \\    
\hline
    &  $8$  &  2.76E-03  &   -    & &  3.02E-03  &   -     & 
            &  3.61E-03  &   -    & &  3.83E-03  &   -   \\
    & $16$  &  1.47E-04  &  4.23  & &  1.50E-04  &   4.33     & 
            &  2.05E-04  &  4.14  & &  2.09E-04  &   4.19\\
    & $32$  &  8.39E-06  &  4.13  & &  8.43E-06  &   4.16     & 
            &  1.24E-05  &  4.05  & &  1.23E-05  &   4.08\\
MHD & $64$  &  5.07E-07  &  4.05  & &  5.08E-07  &  4.05    & 
            &  7.57E-07  &  4.03  & &  7.57E-07  &  4.03 \\        
    & $128$ &  3.13E-08  &  4.02  & &  3.13E-08  &  4.02    & 
            &  4.69E-08  &  4.01  & &  4.69E-08  &  4.01 \\
    & $256$ &  1.94E-09  &  4.01  & &  1.94E-09  &  4.01    & 
            &  2.92E-09  &  4.01  & &  2.91E-09  &  4.00 \\
    & $512$ &  1.21E-10  &  4.00  & &  1.21E-10  &  4.00    & 
            &  1.82E-10  &  4.00  & &  1.82E-10  &  4.00   \\
\hline
\hline
     &  $8$  &  2.27E-02  &   -    & &  2.60E-02  &   -     & 
             &  3.11E-02  &   -    & &  3.48E-02  &   -   \\
     & $16$  &  1.23E-03  &  4.20  & &  1.29E-03  &  4.34   & 
             &  1.75E-03  &  4.15  & &  1.79E-03  &  4.28 \\
     & $32$  &  7.21E-05  &  4.10  & &  7.25E-05  &  4.15   & 
             &  1.04E-04  &  4.07  & &  1.04E-04  &  4.10 \\
RMHD & $64$  &  4.33E-06  &  4.06  & &  4.33E-06  &  4.05    & 
             &  6.45E-06  &  4.01  & &  6.45E-06  &  4.01   \\        
     & $128$ &  2.67E-07  &  4.02  & &  2.67E-07  &  4.02    & 
             &  4.03E-07  &  4.00  & &  4.03E-07  &  4.00   \\
     & $256$ &  1.66E-08  &  4.01  & &  1.66E-08  &  4.01    & 
             &  2.52E-08  &  4.00  & &  2.52E-08  &  3.99   \\
     & $512$ &  1.04E-09  &  4.00  & &  1.04E-09  &  4.00    & 
             &  1.58E-09  &  4.00  & &  1.58E-09  &  4.00    \\
\hline
\end{tabular}
\caption {\footnotesize $L_1$ norm errors and corresponding convergence rates computed for the rotated two-dimensional (columns 2-5) and three-dimensional (columns 6-9) Alfvén wave propagation.
Computations were carried out with the resolution of 8, 16, 32, 64, 128, 256, and 512 grid points along each direction.
}\label{tab:cp_alfven}
\end{table}
\endgroup
Table~\ref{tab:cp_alfven} reports the errors and convergence rates evaluated for a series of 2D and 3D MHD and RMHD computations employing the MP5 and WENOZ reconstructions at increasing resolutions $N_x = \{8, 16, 32, 64, 128, 256, 512\}$.
We use a CFL coefficient of 0.4 and 0.3 for the 2D and 3D case, respectively, and we impose periodic boundary conditions.
The measured convergence rate $O_{L_1}$ well matches the expected value for all the configurations we considered (as shown in the first panel of \refig{fig:cpa_scaling}).
It is interesting to notice that, giving the same configuration, the $L_1$ norm errors are almost identical for the WENOZ or the MP5 schemes.
This confirms that for such a smooth problem, the choice of the high-order spatial reconstruction algorithm does not change results significantly.

In the 3D MHD case, our results may be directly compared to those obtained with the $5^{\rm th}$-order finite difference (FD) scheme of Mignone \cite{Mignone_etal2010} (see Table 1 of that paper), indicating that our FV scheme yields smaller errors for grid sizes lower than $256^3$ zones. 
This difference lessens as the resolution grows (due to the additional order of accuracy of the FD scheme) so that similar errors are eventually found at the resolution of $\sim 256^3$ zones, where $\epsilon_1(B_y) \sim 3.74\cdot10^{-9}$ (for the FD scheme) while $\epsilon_1(B_y) \sim 2.91\cdot10^{-9}$ (with the present method).


For problems without sharp gradients like the one considered here, the employment of a $4^{\rm th}$- (or higher) order scheme leads not only to an increase in accuracy, but also to a saving in computational time for fixed accuracy.
It is therefore instructive to estimate the net gain of using such approach.
Let $\tau$ be the CPU time for updating a single cell with a chosen numerical scheme (e.g. $2^{\rm nd}$ or $4^{\rm th}$-order).
The total CPU time $T$ required to update $N^d$ cells ($d=1,2,3$ indicates the dimensionality of problem) scales as 
\begin{equation}\label{eq:tot_cpu_time}
    T = \tau N^{d+1} \,,\quad 
\end{equation}
where the extra factor $N$ comes from the fact that the hyperbolic time step is inversely proportional to the grid resolution $N$.
Now, let $\epsilon = CN^{-p}$ be the numerical error at the end of the computation, where $p$ is the order of the scheme and $C$ is an unknown constant depending, in general, on the chosen numerical method and the problem at hand.
A $2^{\rm nd}$- and $4^{\rm th}$-order scheme will achieve the same accuracy $\epsilon_2\sim\epsilon_4$ using, respectively, $N_{2}$ and $N_{4}$ grid points related by
\begin{equation}
  N_4 \sim \left(\frac{C_4}{C_2}\right)^{1/4} \sqrt{N_2} \,.
\end{equation}
The previous expression demonstrates that a $4^{\rm th}$-order scheme will require a reduced grid size of about $\sqrt{N_2}$.  
Furthermore, by expressing $N$ as a function of $T$ (from Eq. \ref{eq:tot_cpu_time}) the previous relation can be adjusted in terms of CPU time:
\begin{equation}\label{eq:cpa_time_scaling}
  \frac{T_4}{T_2} = \left(\frac{C_4}{C_2}\right)^{(d+1)/4}
                     \frac{\tau_4}{\tau_2}\frac{1}{\sqrt{N_2^{d+1}}} \,.
\end{equation}
%
From our 3D tests we estimated $\tau_4/\tau_2\sim\{4.3,3.7\}$ and $C_4/C_2\sim\{5.0,8.7\}$ for the MHD and RMHD cases respectively, where we averaged the results produced with the MP5 and WENOZ spatial reconstructions.
Plugging these parameters into \refeq{eq:cpa_time_scaling} we obtain a net gain in computing time, for a given accuracy, that scales as  $T_4/T_2 \sim K/N_2^2$ with $K\sim\{21,32\}$.
Our estimates show, for instance, that for a smooth MHD problem the accuracy obtained with a $2^\mathrm{nd}$-order scheme on a grid with $512\times512\times512$ points could be achieved by our high-order scheme with only $N_4\sim 34$ per direction and a computing time reduced by a factor $\sim 1.2\times10^4$.

Another advantage of using more accurate schemes is the reduction of the numerical dissipation due to the round-off errors introduced by the grid discretization.
We can estimate the diffusion intrinsic to the numerical scheme by looking at the damping of Alfv\'en waves (see \cite{Mignone_2007}).
The decay rate can be expressed as $\Gamma_A = \frac{1}{2}\left(\nu+\eta\right)k^2$, where $\nu$ is the fluid's kinematic viscosity and $\eta$ its resistivity.
We then obtain an approximated value for $\Gamma_A$ by computing the quantity $\Delta\log(\delta B_z)/\Delta t$, where $\delta B_z=\sqrt{\int\int\int B^2_z dxdydz}$ and the difference is evaluated after one period.
From the right panel of \refig{fig:cpa_scaling} we can see how already at very small resolutions ($N\sim8$) the high-order schemes produce an overall numerical dissipation which is more than one order of magnitude lower than the one delivered by the $2^\mathrm{nd}$-order scheme.
Moreover, the gap between the two schemes increases significantly at higher resolutions, as the decay rate scales as $\sim N^{-5}$ in the high-order setup, while we measured a slope of only $\sim-3$ in the low-order case. 

\subsection{Advection of a magnetic field loop} \label{sec:fl}

This problem consists of a weakly magnetized field loop advected by a uniform velocity field. 
Since the total pressure is dominated by the thermal contribution, the magnetic field is essentially transported as a passive scalar and the preservation of the initial circular shape tests the dissipative properties of the scheme \citep{Gardiner_Stone2005, Gardiner_Stone2008, Lee_Deane2009, Mignone_etal2010, Beckwith_Stone2011}.

Following Gardiner \& Stone \citep{Gardiner_Stone2005} and Mignone \citep{Mignone_etal2010}, the computational box is defined by $[-1,1]\times[-0.5,0.5]$ and discretized on $N_x\times N_x/2$ grid cells.  
Density and pressure are initially uniformly set to 1 in the MHD case, while $p=3$ for the RMHD setup (following Beckwith \& Stone \citep{Beckwith_Stone2011}). 
The velocity of the flow is given by
\begin{equation}
    \vec{v} = V_0(\cos\alpha, \sin\alpha) \, ,
\end{equation}
with $V_0 = \sqrt{5}$ in the classical case while $V_0 = \sqrt{5}/10$ for the relativistic case.
The inclination angle is set to $\sin \alpha = 1/\sqrt{5},\, \cos \alpha = 2/\sqrt{5}$.
The magnetic field is defined by means of the vector potential as
\begin{equation}
    A_z = \left\{ \begin{array}{ll}
      A_0(R-r) & \textrm{if} \quad R_1 < r \leq R \,, \\ \noalign{\medskip}
      0   & \textrm{if} \hspace{0.35cm} \textrm{elsewhere} \,,
  \end{array} \right.
\end{equation}
with $A_0 = 10^{-3}$, $R = 0.3$, $R_1 = 0.2R$, $r = \sqrt{x^2+y^2}$.
Simulations have been performed on a computational domain of {$256\times128$} grid points with the LINEAR, WENO3, MP5 and WENOZ schemes up to $t_\mathrm{f}=2$ units for the MHD setup and $t_\mathrm{f}=20$ for the RMHD setup (both corresponding to two advection periods). 
The CFL was set to $0.4$ and double periodic boundary conditions are imposed.

\begin{figure}[!t]
\centering
    \includegraphics[width=0.95\textwidth]{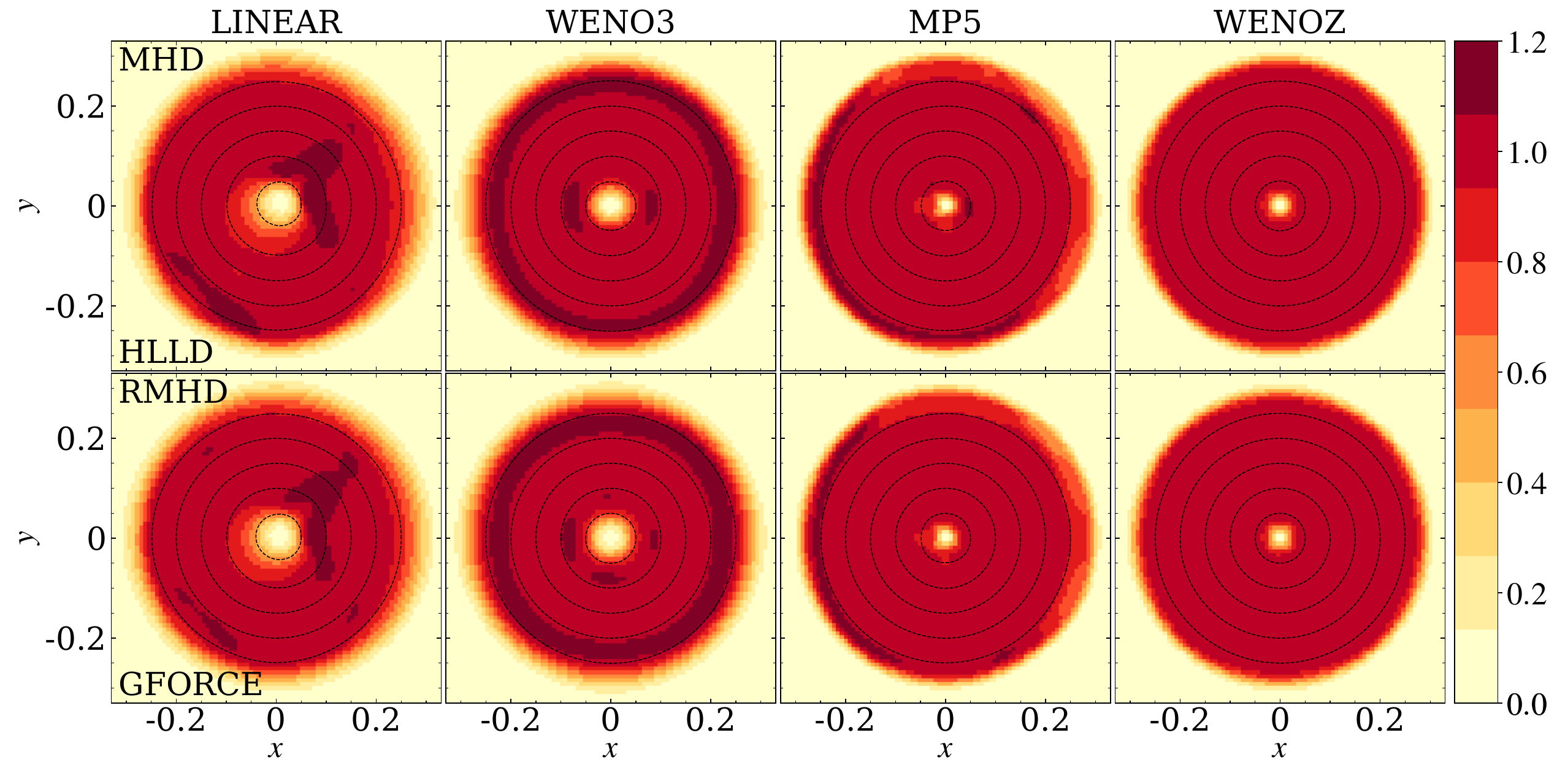}
    \caption{\footnotesize Normalized magnetic energy density for the 2D field loop after two periods at the resolution of $256 \times 128$ grid points in the classical (top) and relativistic (bottom) MHD. 
    Simulations compare LINEAR (leftmost panel), WENO3 (center left panels), MP5 (center right panels) and WENOZ reconstructions (rightmost panels). 
    Magnetic field lines are overlaid as black lines.}
    \label{fig:fieldeq}
\end{figure}

\refig{fig:fieldeq} shows the normalized magnetic energy density together with magnetic field lines with a resolution of $256 \times 128$ grid points for each configuration.
As shown in both sequences of \refig{fig:fieldeq}, smearing and diffusion are significantly reduced as we gradually increase the accuracy from $2^{\rm nd}$ to $4^{\rm th}$-order.
It is interesting to notice that the WENOZ algorithm performs the best in this problem while MP5 has some small amplitude overshoots (undershoots) at the loop tail (head), most likely caused by a small clipping of its limiting procedure.

\begin{figure}[!h]
    \includegraphics[width=0.97\textwidth]{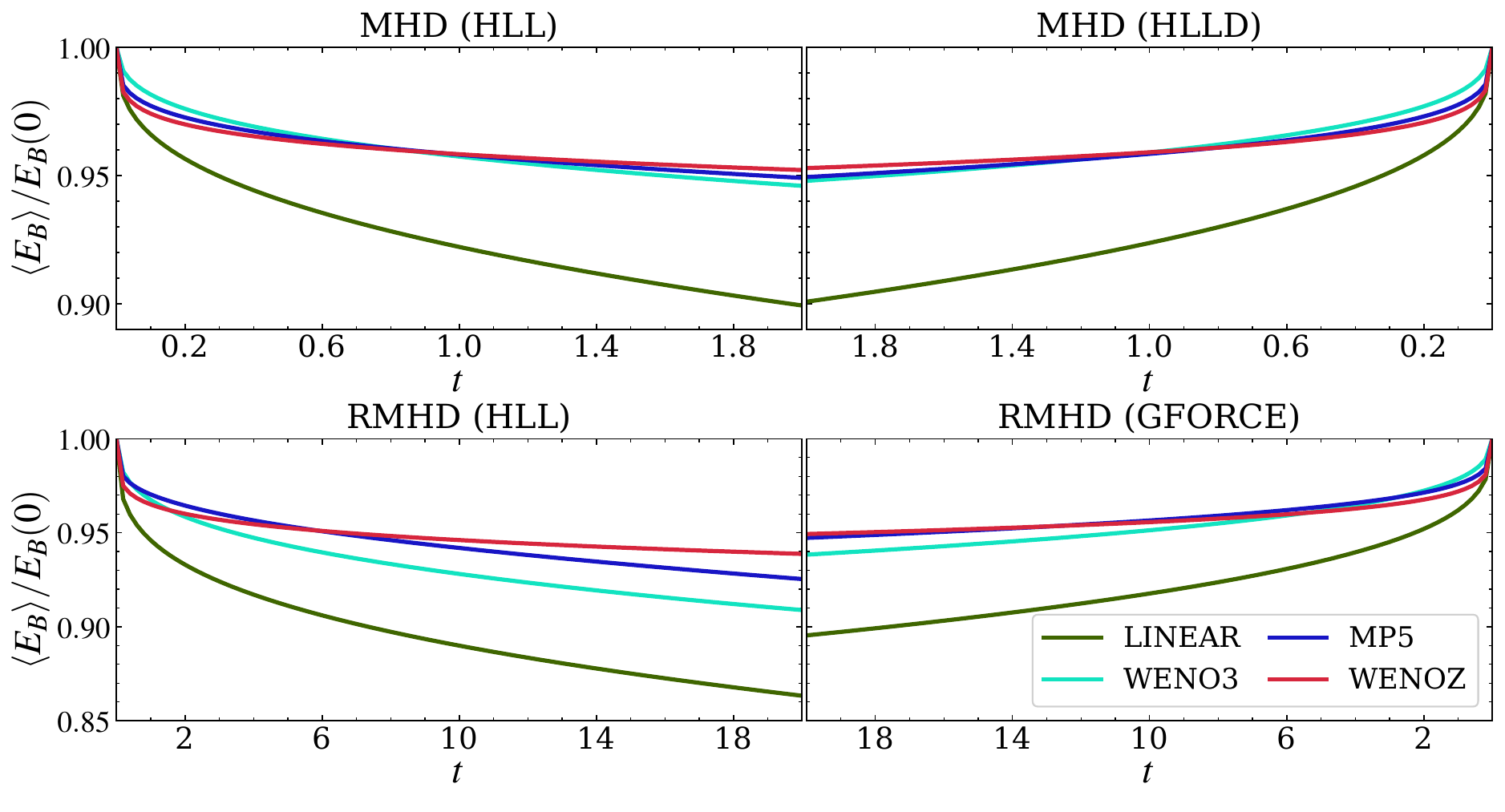}
    \caption{\footnotesize Plot of the normalized magnetic energy density decay of a 2D field loop advected over time at a resolution of $256\times128$ grid points.
    Each simulation has been performed with a $2^{\rm nd}$-order LINEAR (green), a $3^{\rm rd}$-order WENO3+RK3 (cyan), a $4^{\rm th}$-order MP5 (blue), and a $4^{\rm th}$-order WENOZ scheme (red). 
    The top panels show the results obtained in the classical MHD regime, with the HLL (top left panel) and HLLD (top right panel) Riemann solvers.
    Similarly, the bottom panels show the results obtained in the relativistic MHD regime, with the HLL (bottom left panel) and GFORCE (bottom right panel) Riemann solvers.
    Notice that for the results obtained with the HLLD and GFORCE Riemann solvers, the time axes have been reversed.}
    \label{fig:magnenergy}
\end{figure}
Another extremely crucial aspect is the ability of the algorithm to conserve the energy of the system.
\refig{fig:magnenergy} compares the normalized magnetic energy density over time for a set of $4$ simulations employing a $2^{\rm nd}$- (green), a $3^{\rm rd}$- (cyan), and both the $4^{\rm th}$-order accurate (blue and red) schemes, respectively, in the MHD (top panels) and RMHD (bottom panels) regimes.
The top panels compare the results of the HLL + UCT-HLL solver combination with those of the HLLD + UCT-HLLD pair, where the time axis has been reversed to better compare the two sets of simulations.
The RMHD case follows along but with the GFORCE + UCT-GFORCE solver, showing improved results, although the choice of the Riemann solver seems to affect only marginally the conservation of the magnetic energy, with the GFORCE achieving better performance.
The magnetic energy decay is considerably faster for the $2^{\rm nd}$-order scheme compared to the $3^{\rm rd}$- and $4^{\rm th}$-order ones, with the latter yielding the best performance.

Very similar results are also recovered with the FD scheme of Mignone \cite{Mignone_etal2010} (see left panel in Fig. A.9 of that paper) when using either WENOZ or MP5, which dissipate just $\sim 5 \%$ of total magnetic energy at the end of the computation.

\subsection{2D current sheet} \label{sec:cs}

This problem was first introduced by Hawley \& Stone \citep{Hawley_Stone_1995} and also tested by N\'u\~nez-de La Rosa \& Munz both in a MHD \citep{Nunez_2016a} and RMHD \citep{Nunez_2016b} regime (see also references therein).

We use the computational box $[-1,1]\times[-0.5,0.5]$ discretized with $600 \times 600$ grid points.
Density and pressure of an ideal, adiabatic gas with $\Gamma = 5/3$ are initially set to the constant values $\rho = 1$, $p=0.5\beta$, with $\beta$ being the ratio between the thermal and magnetic pressure of the gas.
The magnetic field initially lies in the $y$-direction and it is discontinuous with $B_y = 1$ for $|x| > 0.25$ and $B_y = -1$ otherwise.
We set $v_x = \mathrm{A sin(2\pi y)}, v_y = v_z = 0$, where $A$ is the amplitude of the perturbation.
The parameters $A$ and $\beta$ are both set to the challenging values of $A = \beta = 0.1$ to test the robustness of the underlying scheme.
Simulations have been run up to $t_f = 10$, with a CFL restriction of 0.4 and periodic boundary conditions.

The harmonic velocity perturbation produces non-linear linearly polarized Alfv\'en waves that grow into magnetosonic waves, triggering magnetic reconnection in the regions where the field flips sign.
When $\beta < 1$, namely, in the regime of strongly magnetized fluid, magnetic reconnection forms overpressurized regions that develop magnetosonic waves orthogonal to the field. 
This leads to the conversion from magnetic to thermal energy and to the formation of steep gradients in the regions nearby the field. 
For this reason, this benchmark can be extremely severe in assessing the robustness of the underlying numerical scheme.

\begin{figure}[!h] 
\centering
    \includegraphics[width=0.9\textwidth]{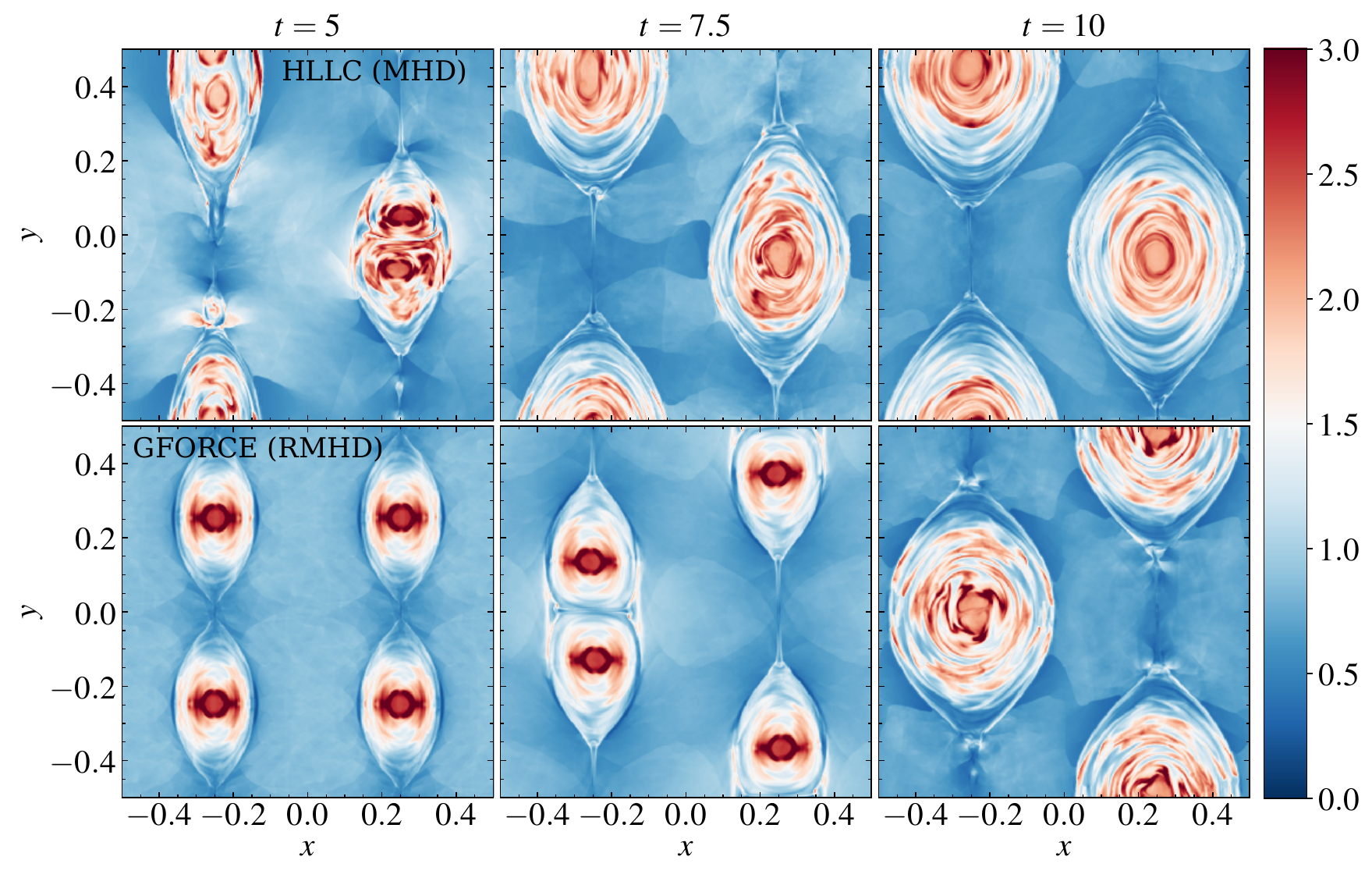}%
    \caption{\footnotesize Coloured maps for the density of the current sheet test problem for the classical (top panel) and relativistic (bottom panel) MHD equations. From left to right snapshots at, respectively, $t = 5, 7.5, 10$ are shown at a resolution of $600\times600$ grid points. Simulations are carried out with the HLLC Riemann solver in the non-relativistic case and with the GFORCE solver in the relativistic case.}
    \label{fig:current_sheets}
\end{figure}

In Fig. \ref{fig:current_sheets} we display density snapshots at $t = 5, 7.5, 10$ for the MHD (top panels) and RMHD setups (bottom panels) obtained, respectively, with the HLLC + UCT-HLL, and with the GFORCE + UCT-GFORCE as solver plus EMF average combination.
This highly non-linear problem tests our novel limiting strategy  where the WENOZ scheme lowers to WENO3 whenever the order reduction procedure is activated, thus increasing the stability range of the scheme.
A similar fallback strategy has been adopted by \citep{Nunez_2016a, Nunez_2016b} who used a simpler Rusanov-Lax Friedrichs solver.

\subsection{Orszag-Tang vortex} \label{sec:ot}

\begin{figure}[!h]
    \centering
    \includegraphics[width=0.9\textwidth]{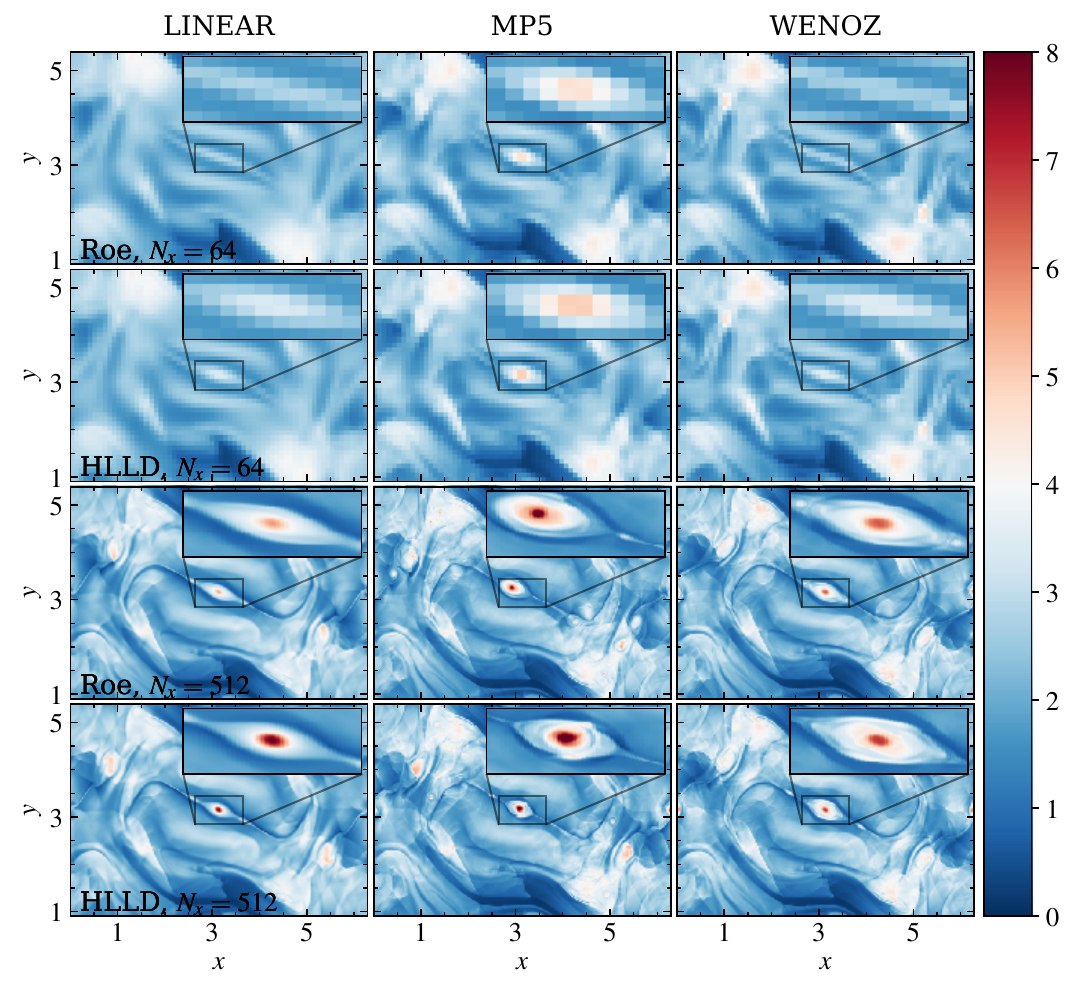}
    \caption{\footnotesize Thermal pressure of the non-relativistic Orszag-Tang vortex computed at $t = 2\pi$ with different algorithms and resolutions. Results are obtained with a $2^{\rm nd}$-order LINEAR (left column), a $4^{\rm th}$-order MP5 (middle column), and a $4^{\rm th}$-order WENOZ scheme (right column).
    Each simulation has been computed with a Roe ($1^{\rm st}$ and $3^{\rm rd}$ rows) and HLLD ($2^{\rm nd}$ and $4^{\rm th}$ rows) Riemann solver with a grid resolution of $N_x = N_y = 64$ ($1^{\rm st}$ and $2^{\rm nd}$ rows) and $N_x = N_y = 512$ ($3^{\rm rd}$ and $4^{\rm th}$ rows).}
    \label{fig:ot_mhd}
\end{figure}

\begin{figure}[!h]
\centering
    \includegraphics[width=0.9\textwidth]{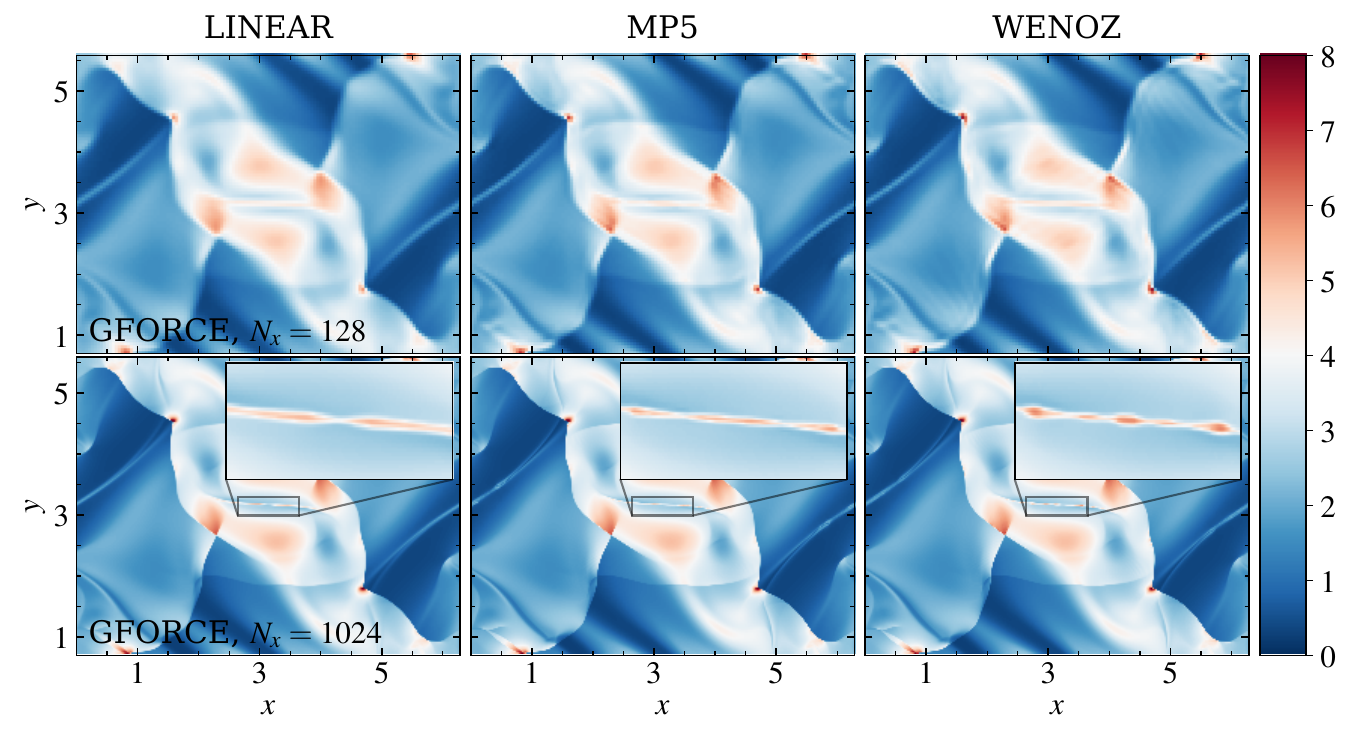}
    \caption{\footnotesize Thermal pressure of the relativistic Orszag-Tang vortex computed at $t = 2\pi$ with different algorithms and resolutions. 
    From left to right, results are obtained with, respectively, a $2^{\rm nd}$-order LINEAR (left column), a $4^{\rm th}$-order MP5 (middle column), and a $4^{\rm th}$-order WENOZ scheme (bottom column). 
    The grid resolution adopted is $N_x = N_y = 128$ (top row) and $N_x = N_y = 1024$ (bottom row). 
     The presence or absence of O-points emerging from the central current sheets (only for the high resolution cases) is highlighted by a zoom on the central region.}
    \label{fig:ot_rmhd}
\end{figure}

The Orszag-Tang MHD vortex \citep{Orszag_Tang_1979} has traditionally been a standard numerical benchmark for inter-scheme comparison.
The test consists of an initial vorticity distribution that spins the fluid counter-clockwise, leading eventually to the formation of a complex structure that includes shocks and turbulence.
The initial density and pressure are set to the constant values $\rho_0 = 25/9$ and $p_0 = 5/3$, while the initial velocity and magnetic fields are $(v_x,v_y) = (-\sin{y}, \sin{x})$ and $(B_x,B_y) = (-\sin{y}, \sin{2x})$, respectively, in the MHD setup, and $(v_x,v_y) = (-0.5\sin{y}, 0.5\sin{x})$ in the RMHD setup (as presented in \cite{Gammie_etal2003,Dumbser_Zanotti2009,Beckwith_Stone2011}).
We ran computations up to $t_f = 2\pi$ with $N_x=N_y=\{64, 512\}$ grid zones for the MHD case, and $N_x=N_y=\{128, 1024\}$ for the RMHD case using periodic boundary conditions and a CFL number of $0.5$.
We repeated the computations by choosing two different Riemann solvers in the MHD case (Roe and HLLD, always with the UCT-HLLD average), and only the GFORCE Riemann solver coupled to the UCT-GFORCE EMF averaging algorithm in the relativistic case.
Likewise, we also compared 3 different reconstruction schemes, namely, LINEAR, MP5, and WENOZ.

Multiple shock-vortex interactions regulate the dynamics of the system, leading to the formation of an inclined current sheet at the center of the domain. 
The magnetic energy is gradually dissipated and the current sheet twists, leading to the structures observed in the thermal pressure distribution shown in \refig{fig:ot_mhd}.
For the MHD setup, even at modest resolutions (64 points) the $4^{\rm th}$-order scheme with MP5 reconstruction and the HLLD Riemann solver allows to distinguish at the center of the computational domain the formation of a magnetic island (an O-point). 
It is well known (see \citep{Landi_etal2015, Papini_etal2019}) that the tearing instability is expected to develop on the ideal (Alfv{\'e}nic) timescales of an ideal MHD simulation only if the numerical dissipation is sufficiently low.
At higher resolutions, all simulations produce high-pressure regions, albeit small-scale structures are best appreciated with the high-order runs.

Similarly to Mignone \& Del Zanna \cite{Mignone_DelZanna2021}, we can assess more quantitatively the improvements introduced by our $4^{\rm th}$-order scheme by computing the parameter $\mu_p=\mathrm{max}(p)/\langle p\rangle$, i.e. the pressure maximum normalized by the average pressure at the end of a run.
As this quantity generally increases with better resolution, we use it as a proxy to assess the decrease of numerical dissipation.
For instance, the values of $\mu_p$ obtained with the MP5 reconstruction and HLLD solver against those obtained with the LINEAR one yield an average increase in effective resolution of $\sim 2.23$, which leads to an average speedup of $\sim 2.58$ (taking into account the extra computational cost of the $4^{\rm th}$-order scheme estimated in \S 5.3).
This is, however, only a rough estimate that cannot fully characterize the impact of our scheme on the typical numerical dissipation length-scale in a turbulent plasma, for which more dedicated tests and analysis are required.

For the RMHD case, the final stage of the simulation produces a central low-pressure cavity with a thin high-pressure filament, which exhibits finer structures for less dissipative schemes.
While at a resolution of $128$ points there are no apparent deviations between the different setups, at higher resolution ($1024$ points) the numerical resistivity from the $4^{\rm th}$-order scheme is low enough to lead to the production of multiple O-points in the high-pressure filament, although they remain still quite unresolved. 
Our results confirm the overall robustness of the scheme as well as its ability to capture small-scale structures that would otherwise be lost with a lower order scheme.

\subsection{3D Blast Wave} \label{sec:blast}

\begin{table}
\centering
\begin{tabular}{l c c c c c c c c c c c} 
\hline
Case & $\rho_{\rm in}$ & $\rho_{\rm out}$  &  $p_{\rm in}$ & $p_{\rm out}$ & $B_0$ & $\Gamma$ & $r_0$ & $L$ & $t_{\rm stop}$ & Solver & En. Corr.
\\ \noalign{\medskip}
\hline
C1 (MHD)  &   $1$    &      $1$      &      $10$    &    $0.1$      & $3$         &  $5/3$   & $0.1$ & $1$ &      $0.2$     & HLL   &   NO 
\\ \noalign{\medskip} 
C2 (MHD)  &$1$  &      $1$         &    $10^3$  &    $0.1$      &$100/\sqrt{4\pi}$     &  $1.4$   & $0.1$ & $1$ &      $0.01$    & HLL   &   YES
\\ \noalign{\medskip}
R1 (RMHD)  &$10^{-2}$  &      $10^{-4}$   &      $1$   & $5\times10^{-3}$ & $1$     &  $4/3$   & $0.08$ & $6$&      $4$    & GFORCE &  YES
\\ \hline  
\end{tabular}
\caption{\footnotesize Parameters used in the 3D blast wave problem.
 }
\label{tab:blast3D}
\end{table}
 
We finally consider the 3D magnetized blast wave problem in order to assess the robustness of the algorithms under strong magnetization conditions.
Despite its simplicity, the blast wave problem is a particularly effective benchmark in testing the solver's ability at handling MHD wave degeneracies parallel and perpendicularly to the field orientation.

The initial condition consists of a uniform medium with density and pressure values set, respectively, to $\rho_{\rm out}$ and $p_{\rm out}$ and threaded by an oblique constant magnetic field,
\begin{equation}
   \vec{B}=B_0\Big(\sin{\theta}\cos{\phi},\,
                   \sin{\theta}\sin{\phi},\,
                   \cos{\theta}\Big)
\end{equation}
where $\theta = \pi/2$ and $\phi=\pi/4$ are used in what follows.
Note also that oblique configurations make the computation considerably more challenging than in the grid-aligned cases (see, for instance the discussion in \S 4.4 of Mattia \& Mignone \cite{Mattia_Mignone2022}).
The computational domain is a cube of length $L$ centered at the origin with zero-gradient boundary conditions.
An over-pressurized spherical region where pressure and density can have different values, namely $\rho_{\rm in}$ and $p_{\rm in}$, is initially set for $r < r_0$.  

We examine three different cases with parameters listed in table \ref{tab:blast3D}. 
Cases C1 is taken from Felker \& Stone \cite{Felker_Stone2018}, with the exception that we consider a magnetization three times as large here.
Case C2 proposes a more severe configuration as initially presented by Balsara \& Spicer \cite{Balsara_Spicer1999} (see also \cite{Mignone_DelZanna2021}).
Lastly, in case R1 (from \cite{Mattia_Mignone2022}) we apply our $4^{\rm th}$-order method to the solution of the relativistic MHD equations.
Note that the energy correction step (see \cite{Balsara_Spicer1999} and, e.g., see Sec. 6.4 of \cite{Mignone_DelZanna2021}) is applied to both cases C2 and R1, while case C1 has the largest magnetization our scheme can afford without it.
This correction step is still performed at $2^{\rm nd}$-order level and it was found to substantially improve the robustness of the code for larger magnetic field strengths.
Computations are performed using $192\times192\times192$ grid zones.

\begin{figure}[!h]
    \includegraphics[width=0.97\textwidth]{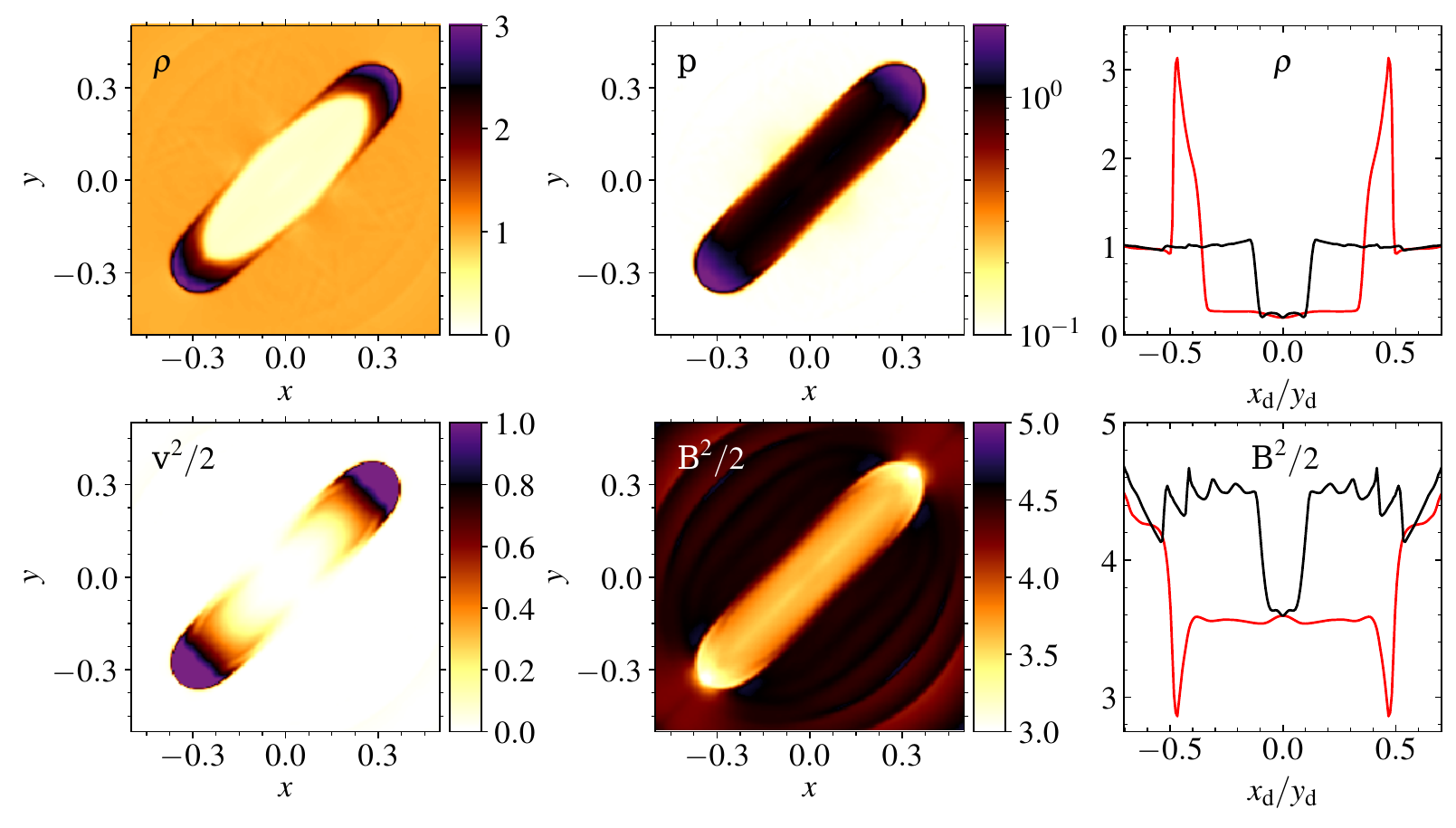}
    \caption{\footnotesize MHD 3D blast wave problem (case C1) at $t=0.2$.
    The left and central panels show density (top left), pressure (top middle, in logarithmic scale), specific kinetic energy (bottom left) and magnetic energy (bottom right) at $z = 0$ for a simulation employing $192\times192\times192$ grid points.
    In the rightmost panels we show the density (top) and magnetic energy density (bottom) on the major (red line) and minor (black line) diagonals.
    }
    \label{fig:blast_c1}
\end{figure}

\begin{figure}[!h]
    \includegraphics[width=0.97\textwidth]{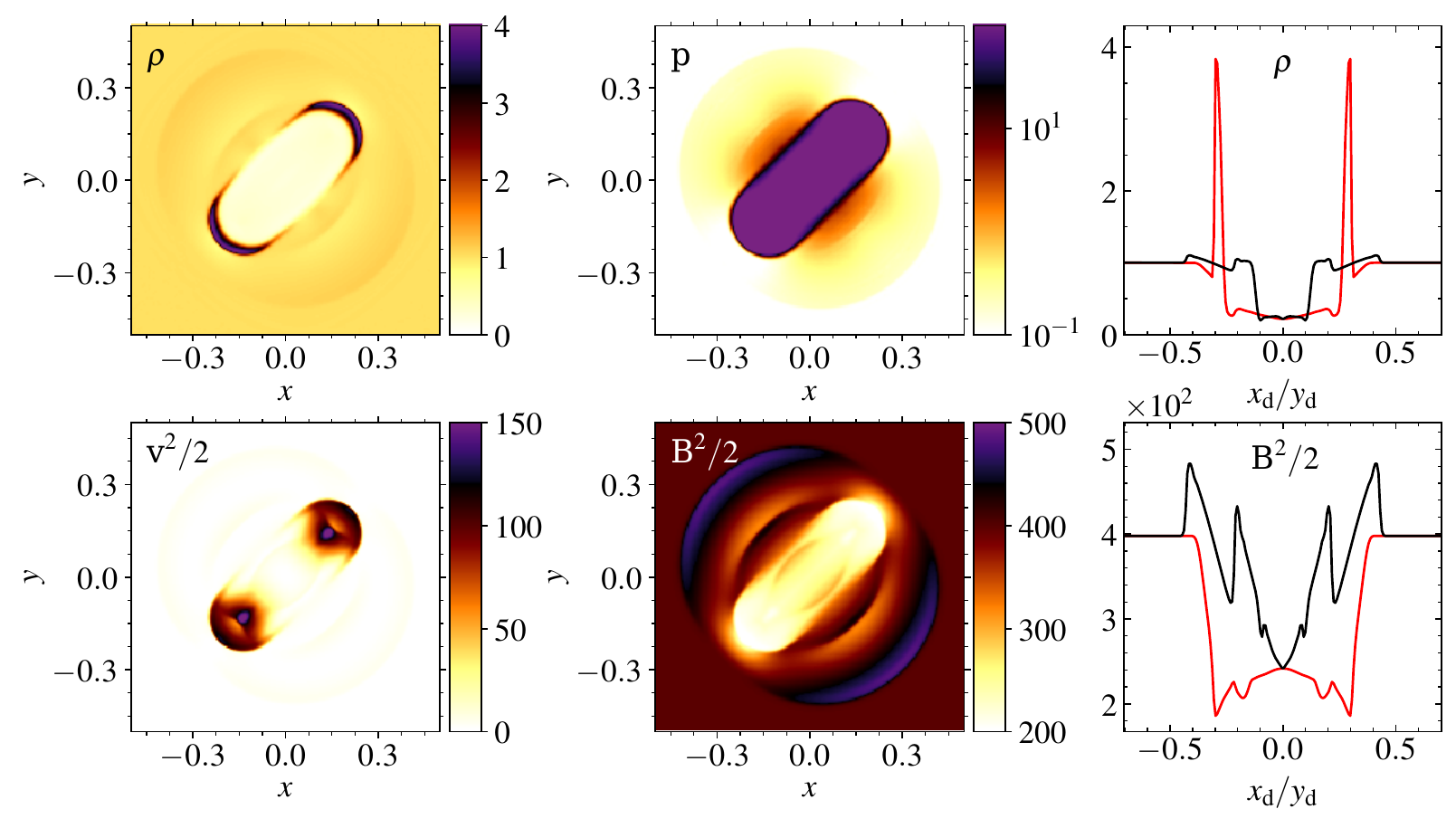}
    \caption{\footnotesize Same as Fig. \ref{fig:blast_c1} but for the case C2 at $t = 0.01$.}
    \label{fig:blast_c2}
\end{figure}

\begin{figure}[!h]
    \includegraphics[width=0.97\textwidth]{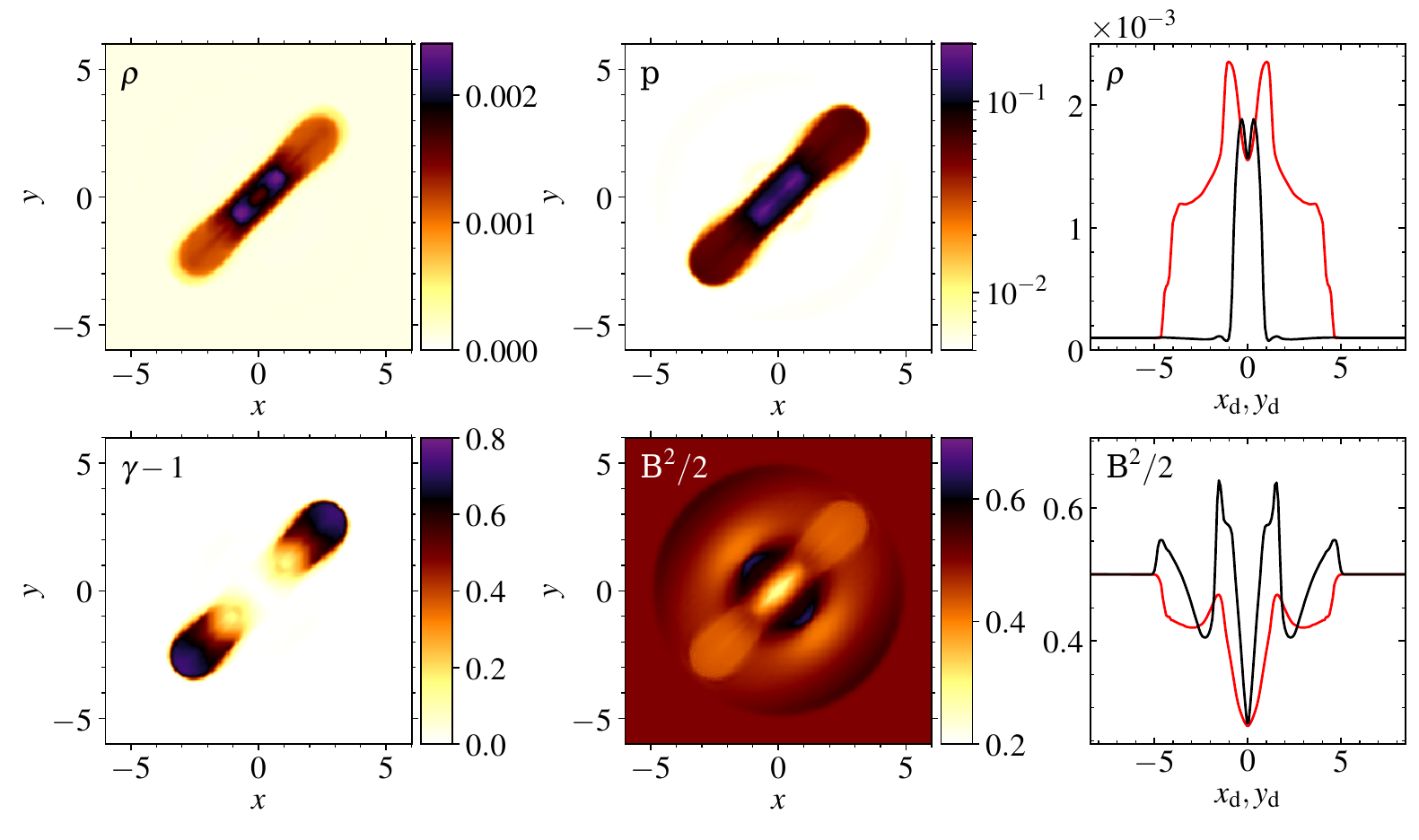}
    \caption{\footnotesize Same as Fig. \ref{fig:blast_c1} but for the case R1 at $t = 4$. In the bottom left panel, the specific kinetic energy has been replaced by the relativistic counterpart.}
    \label{fig:blast_r1}
\end{figure}

Results are shown in Fig. \ref{fig:blast_c1} (C1), \ref{fig:blast_c2} (C2) and \ref{fig:blast_r1} (R1) for the three selected cases.
The explosion spawns an oval-shaped structure delimited by an outer fast forward shock and the presence of a magnetic field makes the propagation highly anisotropic by compressing the gas in the direction parallel to the field. 
Gas motion takes place along the field lines generating two oppositely moving plasma blobs. 
In the perpendicular direction, the outer fast shock becomes magnetically dominated with very weak compression. 
Discontinuous waves and smooth structures are well resolved by our scheme and the point-symmetry is preserved throughout the evolution.
We point out that the fallback approach played an essential role for this test thus strengthening the benefits of a low/high order hybrid scheme while resolving complex flows featuring both smooth as well as discontinuous solutions.

\subsection{3D Cloud-Shock Interaction} \label{sec:shock_cloud}

We propose a 3D version of the cloud-shock interaction test previously introduced by \citep{Balsara_2001, Ziegler_2004, Miniati_2011, Helzel_etal2011} in its MHD version, and by Mignone \citep{Mignone_PLUTO2012} for the RMHD case.
For both configurations, the initial condition consists of a plane discontinuity propagating a fast shock and a rotational wave. 
Specifically, in the MHD case, the initial state $\cV = (\rho, v_{x}, v_{y}, v_{z}, B_{x}, B_{y}, B_{z}, p)$ is initialized as
\begin{equation}\label{eq:mhd_shock_cloud}
  \left\{\begin{array}{lclr} 
    \cV_L &=&\DS \left(3.86859,\,0,\,0,\,0,\, 
                        0,\, 2.1826182,\, -2.1826182,\, 167.345 \right)
          & \quad{\rm for}\quad x < 0.6\, ,
    \\ \noalign{\medskip}
    \cV_R &=&\DS \left(1, \,-11.2536,\, 0,\,  0,\, 0,\, 0.56418958,\, 0.56418958,\, 1 \right)
          & {\rm for}\quad x > 0.6\, ,
  \end{array}\right.
\end{equation}
while in the RMHD test as
\begin{equation}\label{eq:rmhd_shock_cloud}
  \left\{\begin{array}{lclr} 
    \cV_L &=&\DS \left(39.5052,\,0,\,0,\,0,\, 
                        0,\, 0,\, 1.9753,\, 129.72386 \right)
          & \quad{\rm for}\quad x < 0.6\, ,
    \\ \noalign{\medskip}
    \cV_R &=&\DS (1, \,-\sqrt{0.99},\, 0,\,  0,\, 0,\, 0,\, 0.5,\, 10^{-3})
          & {\rm for}\quad x > 0.6\, .
  \end{array}\right.
\end{equation}
At $\Vec{x} = (0.8, 0.5, 0.5)$  an overdense ($\rho = 10$) spherical cloud of radius $0.15$ is initially in hydrostatic equilibrium with the ambient plasma under the adiabatic equation of state with $\Gamma = 5/3$ ($\Gamma = 4/3$ in the special relativistic case).
For both setups the computational domain consists of the unit cube, with outflow (zero-gradient) boundary conditions employed everywhere, except at the lower $x$- boundary, where the initial condition is imposed.
 
\begin{figure}[!h]
    \centering
    \includegraphics[width=0.9\textwidth]{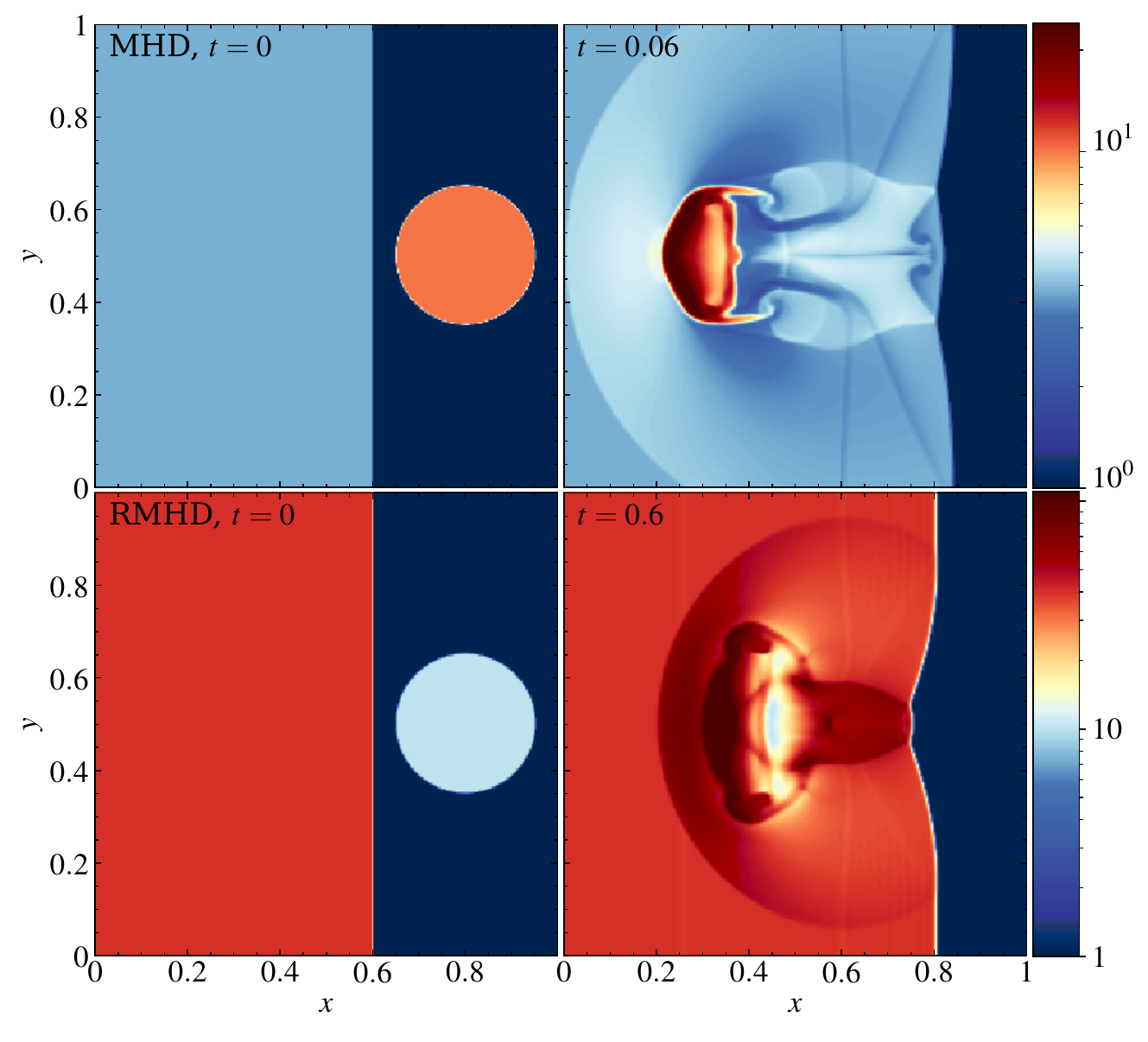}
   
    \caption{\footnotesize Non-relativistic (top panels) and relativistic (bottom panels) cloud-shock interaction at initial ($t = 0$, left panels) and final time ($t = 0.06$ for the non-relativistic case, and $t = 0.6$ for the relativistic case, right panels). Shown is the density at $z = 0.5$ in logarithmic scale. Both simulations ran with $256\times256\times256$ grid points. The colorscales span over a range of $[1,24]$ for the non-relativistic case and $[1,110]$ for the relativistic case. }
    \label{fig:sc_test}
\end{figure}

During the evolution, the magnetic shock propagates to the right, and the high density cloud moves supersonically into the shock front.
At the final evolution time ($t_f = 0.06$ MHD, $t_f = 0.6$ RMHD) the cloud has been completely shocked and it has developed its characteristic mushroom-like structure.
\refig{fig:sc_test} shows the initial and final density condition for a $256\times256\times256$ grid points simulation.
The $4^{\rm th}$-order scheme properly recovers the expected structures.
It is important to mention that, for this particular test problem, the employment of the order reduction procedure extended to the selective fluxes integration has been fundamental for the correct evolution of the simulations.

\section{Conclusions}
\label{sec:summary}
%
%
%
%

In this paper we have presented a genuinely $4^{\rm th}$-order accurate finite volume method for the solution of the classical and special relativistic MHD equations employing the constrained transport algorithm.
Our scheme has been implemented in the PLUTO code for astrophysical plasma dynamics \citep{Mignone_PLUTO2007} and is rooted over the method originally proposed by McCorquodale \& Colella \citep{Corquodale_Colella2011}, adding a number of innovative aspects that yield an accurate, robust and efficient computational tool.
In fact, our scheme maximizes the benefits of the FV formalism (e.g., robustness when dealing with shocks, discontinuities, and steep gradients) along with techniques inspired by the FD formalism, such as the introduction of pointwise to pointwise reconstruction operations that ease up the structure of the scheme.
To this aim, we revisited standard spatial reconstruction schemes (such as WENOZ and MP5), traditionally based on the knowledge of 1D volume averages, so as to provide $4^{\rm th}$-order accurate left and right states at zone interfaces by directly interpolating point values rather than volume averages.
Pointwise reconstructions can be accomplished in primitive or characteristic variables and they lead to a compact stencil and a more cost-effective scheme.
Moreover, our implementation is endowed with the generalized upwind constrained transport (UCT) of Mignone \& Del Zanna \citep{Mignone_DelZanna2021}, plus a new generalization of the UCT-GFORCE averaging scheme for special relativistic MHD that has been widely validated in the presented benchmarks.

The performance of the method has been assessed against selected 2D and 3D numerical test problems, demonstrating: i) genuine $4^{\rm th}$-order accuracy in smooth problems, ii) reduced numerical dissipation iii) the code's ability to correctly capture propagating discontinuities and to resolve small-scale structures and turbulence.
The robustness of the method is reinforced by a fallback approach that reverts the spatial order of the scheme to $2$ in proximity of steep gradients.
To identify such critical regions, we have proposed a new discontinuity detector - based on a derivatives ratio - which employs up to the fourth derivative to pinpoint high-frequency oscillations typically arising in proximity of a discontinuity.
The order reduction procedure turned out to be, in our experience, a crucial element when undertaking severe configurations, allowing our high-order scheme to successfully carry out computations otherwise feasible only with $2^{\rm nd}$-order schemes (see, e.g., configuration C2 \S\ref{sec:blast}). 

We point out that, despite the augmented complexity of high-order algorithms, genuine $4^{\rm th}$-order accuracy in smooth problems leads to a net saving of computational time.
In \S\ref{sec:cp} we demonstrated that, for a given accuracy, the high-order scheme will require, approximately $\sqrt{N_2}$ grid cells when compared to a $2^{\rm nd}$-order scheme with a corresponding reduction of CPU time that scales as $\sim N_2^{-(d+1)/2}$, where $d$ is the number of spatial dimensions.
When applied to simple 3D computations, for instance, we found that the accuracy reached by a $2^\mathrm{nd}$-order scheme with $512^3$ grid points is equally obtained when employing our $4^{\rm th}$-order scheme with only $\sim 34^3$ points, with a corresponding reduction of the total computational time by a few orders of magnitude.

Better accuracy leads also to reduced numerical dissipation and thus an enhanced potential at describing spatial scales that would otherwise be challenging to probe within a traditional $2^{\rm nd}$-order framework.
Our results indicate that the decay rate of numerical diffusion is $\sim N^{-5}$ for our novel scheme, and only $\sim N^{-3}$ for the traditional scheme.
For all these reasons our $4^{\rm th}$-order scheme proved to be a computational tool suited for the next generation of numerical simulations in computational fluid dynamics.
In a companion paper we will extend this formalism to non-Cartesian geometries.

Compared to other high-order competitive methods, such as discontinuous Galerkin (DG), our FV scheme requires fewer equations to be evolved in time (e.g. $8$ instead of $32$ for a $4^{\rm th}$-order method), does not reduce stability with increasing order and it is better suited for flows containing smooth and discontinuous features.
Its parallel implementation requires, however, several more inter-processor communications (one per Runge-Kutta stage). 
A more thorough comparison, though, has to be considered in a separate paper.


\vspace*{2ex}\par\noindent
{\bf Acknowledgments.}
We thank L. Del Zanna for useful discussions and comments.
The PLUTO code is publicly available and the simulation data will be shared on reasonable request to the corresponding author.

\vspace{7pt}\noindent
This project has received funding from the European Union's Horizon Europe research and innovation programme under the Marie Sk\l{}odowska-Curie grant agreement No 101064953 (GR-PLUTO).

\vspace{7pt}\noindent
This work has received funding from the European High Performance Computing Joint Undertaking (JU) and Belgium, Czech Republic, France, Germany, Greece, Italy, Norway, and Spain under grant agreement No 101093441 (SPACE).

\vspace{7pt}\noindent
This paper is supported by the Fondazione ICSC, Spoke 3 Astrophysics and Cosmos Observations. 
National Recovery and Resilience Plan (Piano Nazionale di Ripresa e Resilienza, PNRR) Project ID CN\_00000013 \quotes{Italian Research Center on High-Performance Computing, Big Data and Quantum Computing}  funded by MUR Missione 4 Componente 2 Investimento 1.4: Potenziamento strutture di ricerca e creazione di \quotes{campioni nazionali di $R\&S$ (M4C2-19 )} - Next Generation EU (NGEU).

\vspace{7pt}\noindent
We also acknowledge the OCCAM supercomputing facility available at the Competence Centre for Scientific Computing at the University of Torino.

\appendix

\section{Laplacian Operators in Cartesian Coordinates}\label{app:laplacians}
%

We start by demonstrating the validity of \refeq{eq:v2p} and deriving the expression of the $2^{\rm nd}$-order accurate Laplacian operator for a given function $U(x)$ in a Cartesian mesh in one dimension. 
The possibility to straightforwardly generalize the treatment to multiple dimensions is guaranteed when considering Cartesian meshes due to the orthogonality of the standard basis.
The (1D) cell-average is given by
\begin{equation}
  \DS \av{U}_{\cc} = \frac{1}{h} \int_{-\frac{h}{2}}^{\frac{h}{2}} U(x) \,dx \, ,
\end{equation}
where $h$ is the cell width.
By expanding in Taylor series $U(x)$ near the cell-center $x_c$ up to $4^{\rm th}$-order we obtain
\begin{equation}
  \DS \av{U}_{\cc} = \frac{1}{h} \int_{-\frac{h}{2}}^{\frac{h}{2}} 
       \left[U(x_c) + U'(x_c)(x-x_c) + \frac{U''(x_c)}{2}(x-x_c)^2 + \frac{U'''(x_c)}{6}(x-x_c)^3 + O(x^4)\right] \,dx \, ,
\end{equation}
and, since odd integrands vanish when integrated over symmetrical domains, from straightforward calculation we get
\begin{equation}
  \DS \av{U}_{\cc} = U(x_c) + \frac{h^2}{24}U''(x_c) + O(h^4) \, . \label{eq:uc}
\end{equation}
We now obtain a $2^{\rm nd}$-order accurate expression of the $2^{\rm nd}$-order derivative $U''$.
By expanding in Taylor series up to $4^{\rm th}$-order the nearby average values: $\av{U}_{\cc-\hvec{e}_x}$, $\av{U}_{\cc+\hvec{e}_x}$ we get
\begin{equation}
 \begin{array}{lc}
    \DS \av{U}_{\cc-\hvec{e}_x} = \frac{1}{h} \int_{-\frac{3h}{2}}^{-\frac{h}{2}} 
        \left[U(x_c) + U'(x_c)(x-x_c) + \frac{U''(x_c)}{2}(x-x_c)^2 + \frac{U'''(x_c)}{6}(x-x_c)^3 + O(x^4)\right] \,dx \, ,
        \\ \noalign{\medskip}
    \DS \av{U}_{\cc+\hvec{e}_x} = \frac{1}{h} \int_{\frac{h}{2}}^{\frac{3h}{2}} 
        \left[U(x_c) + U'(x_c)(x-x_c) + \frac{U''(x_c)}{2}(x-x_c)^2 + \frac{U'''(x_c)}{6}(x-x_c)^3 + O(x^4)\right] \,dx \, .
 \end{array}
\end{equation}
Since the interval is no longer symmetric, the odd powers no longer integrate to zero.
The final result is
\begin{align}
   \av{U}_{\cc-\hvec{e}_x}&=U(x_c) - U'(x_c)h + \frac{13}{24}U''(x_c)h^2 - \frac{5}{24}U'''(x_c)h^3 + O(h^4) \label{eq:uc-e} \, ,
   \\
   \av{U}_{\cc+\hvec{e}_x}&=U(x_c) + U'(x_c)h + \frac{13}{24}U''(x_c)h^2 + \frac{5}{24}U'''(x_c)h^3 + O(h^4) \label{eq:uc+e} \, .
\end{align}
Adding \refeq{eq:uc-e} to \refeq{eq:uc+e} and subtracting twice \refeq{eq:uc} 
\begin{equation}
    \DS \av{U}_{\cc+\hvec{e}_x} + \av{U}_{\cc-\hvec{e}_x} - 2 \av{U}_{\cc} = U''h^2 + O(h^4) \, ,
\end{equation}
we can express the $2^{\rm nd}$-order accurate second derivative in terms of volume-averaged quantities as
\begin{equation} \label{eg:lap}
    \DS U'' = \frac{\av{U}_{\cc+\hvec{e}_x} - 2 \av{U}_{\cc} + \av{U}_{\cc-\hvec{e}_x} }{h^2} + O(h^2) \, ,
\end{equation}
which defines the Laplacian operator in \refeq{eq:v2p}.

Conversely, \refeq{eq:p2v} can be formally obtained applying a Simpson's quadrature rule by integrating over the cell width $h$
\begin{equation} \label{eq:simpson}
    \DS \int_{-\frac{h}{2}}^{\frac{h}{2}} U(x)\,dx \simeq  
    \DS \int_{-\frac{h}{2}}^{\frac{h}{2}} 
        \left(\alpha x^2+\beta x+ \gamma \right)\,dx \, .
\end{equation}
The function $U(x)$ is approximated by a $2^{\rm nd}$-order polynomial in the coefficients $\alpha , \beta , \gamma$. 
The polynomial in the right hand side of \refeq{eq:simpson} is the best approximation of $U(x)$ when the coefficients $\alpha , \beta , \gamma$ are set by evaluating the function in the $3$ nearby cell centers:
\begin{equation}
    \DS \alpha = \DS \frac{U_{\cc+\hvec{e}_x} - 2 U_{\cc} + U_{\cc-\hvec{e}_x}}{2} \, , \quad
    \DS \beta = \DS \frac{U_{\cc+\hvec{e}_x} - U_{\cc-\hvec{e}_x}}{2} \, , \quad
    \DS \gamma = \DS U_{\cc} \, .
\end{equation}
By replacing these expressions in \refeq{eq:simpson} and by direct calculation of the integral we obtain \refeq{eq:p2v}.

\section{Explicit Expressions for the UCT Scheme} 
\label{app:uct_averages}
%
%

Here, for the sake of completeness and to also amend some equations containing typos, we briefly recap the most relevant expressions needed in our UCT scheme, as originally presented by Mignone \& Del Zanna \cite{Mignone_DelZanna2021}.
The edge-centered EMF (Eq. \ref{eq:emf}) requires the reconstructed values of the transverse velocities as well as the linear combination coefficients $a$ and $d$ (for the centred and diffusive flux contribution, respectively).

\begin{itemize}[leftmargin=*]
    
\item \textit{Transverse Velocities}: when using the HLL, HLLC or HLLD classical MHD Riemann solvers at a zone interface, the transverse velocities are obtained, e.g., at an $x$-face, as
\begin{equation}\label{eq:vt}
  \overline{v}_{t,\xf} = \frac{  \alpha_x^R v^L_{t,\xf}
                               + \alpha_x^L v^R_{t,\xf}}
                              {\alpha^R_x + \alpha^L_x} \,, \quad (t=y,z) \,,
\end{equation}
where $v^{L/R}_{t,\xf}$ are the left / right transverse velocities reconstructed at an $x$-face, 
\begin{equation}\label{eq:alphaHLL}
    \alpha_x^R =  \max(0, \lambda^R_{\xf}) ,\quad
    \alpha_x^L = -\min(0, \lambda^L_{\xf})
\end{equation}
with $\lambda^{L/R}_{\xf}$ being the smallest and largest characteristic speeds (see Eq. 14 of \cite{Mignone_DelZanna2021}). 
In the case of the GFORCE Riemann solver instead, we use
\begin{equation}\label{eq:UCT_GFORCE_vt}
  \overline{v}_t =
  \omega\left[v_t^{\rm LW} - \frac{\tau_x}{2}\left(v^R_t - v_t^L\right) v_x^{\rm LW}\right]
  + (1-\omega) \left(\frac{v_t^L + v_t^R}{2}\right)
\end{equation}
(note that Eq. 51 of \cite{Mignone_DelZanna2021} contains \quotes{$y$} instead of \quotes{$t$}), with $\omega = 1/(1+c)$ ($c$ is the Courant factor, $c = 1$ yields the FORCE scheme), $\vec{v}^{\rm LW}$ is the velocity in the Lax-Wendroff state (see. Eq. 48 of \cite{Mignone_DelZanna2021}), and $\tau_x = 1/\max(|\lambda^L_{\xf}|,\, |\lambda^R_{\xf}|)$.

\item \textit{Combination Coefficients $a$ and $d$}.
For the HLL or HLLC Riemann solvers (away from degeneracies), the combination coefficients are evaluated as arithmetic averages across two adjacent faces (Eq. 34 and 35 of \cite{Mignone_DelZanna2021}) using the values available with the 1D Riemann solver: 
\begin{equation}\label{eq:HLLcoeffs}
  a^L_x = \frac{\alpha_x^R}{\alpha^R_x + \alpha^L_x}\,,\quad   
  a^R_x = \frac{\alpha_x^L}{\alpha^R_x +\alpha^L_x}\,,\quad
  d^L_x = d^R_x = \frac{\alpha^R_x\alpha^L_x}{\alpha^R_x + \alpha^L_x} \,,
\end{equation}
where $\alpha_x^{L/R}$ have been given in Eq. (\ref{eq:alphaHLL}).
In the limit $B_x\to0$, the corresponding expressions for the HLLC solver are given by Eq. 38 of \cite{Mignone_DelZanna2021}.
For the HLLD solver, instead, we employ
\begin{equation}\label{eq:UCT_HLLD_ad}
  a^L = \frac{1 + \nu^*}{2}\,,\quad
  a^R = \frac{1 - \nu^*}{2}\,,\quad
  d^s =   \frac{1}{2}(\nu^s - \nu^*)\tilde{\chi}^s
        + \frac{1}{2}\left(|\lambda^{*s}| - \nu^*\lambda^{*s}\right) \,,
        \quad ({\rm s} = L,R)
\end{equation}
where $\tilde{\chi}^s = (\lambda^{*s} - \lambda^s)\chi^s$, while
\begin{equation}\label{eq:UCT_HLLD_nu}
  \nu^s = \frac{\lambda^{*s} +\lambda^s}{|\lambda^{*s}| + |\lambda^s|} \,,\qquad
  \nu^* = \frac{\lambda^{*R} + \lambda^{*L}}{|\lambda^{*R}| + |\lambda^{*L}|}\,,
\end{equation}
and $\lambda^{*L/R}$ are the Alfv\'en velocities of the HLLD fan (see Eq. 40 of \cite{Mignone_DelZanna2021}).

\end{itemize}

\bibliographystyle{elsarticle-num}
\bibliography{paper}

%
%
\end{document}